\documentclass[aps,pra,showpacs,twoside,twocolumn,10pt]{revtex4-2}
\usepackage[colorlinks=true, citecolor=blue, urlcolor=blue ]{hyperref}
\usepackage{epsfig,newlfont,amssymb,amsfonts,amsmath,bm,subfigure,palatino,mathtools,amsthm,braket,times,soul,enumitem,color}
\usepackage[normalem]{ulem}

\usepackage{tikz}
\usetikzlibrary{plotmarks}
\usetikzlibrary{arrows.meta, positioning, decorations.pathreplacing}
\usepackage{textcomp}

\usepackage{blindtext}
\usepackage{graphicx}
\usepackage{amsmath}
\usepackage{bm}
\usepackage{hyperref}
\usepackage{geometry}
\usepackage{amsthm}
\usepackage{gensymb}
\usepackage{physics}
\newcommand{\ab}{\color{black}}

\begin{document}
\title{Reduced-State Stabilizer R\'enyi Entropy as a Probe of Quantum Criticality in the Transverse ANNNI Model and the Quantum Compass Model}
\author{Santanu Sarkar\(^{1, \dagger}\), George Biswas\(^{2, \dagger, \star}\), Jun-Yi Wu\(^{2,3,4}\), and Anindya Biswas\(^{1, \star}\)}

\affiliation{\(^1\) Department of Physics, National Institute of Technology Sikkim, Ravangla, Namchi, Sikkim 737139, India\\ \({^2}\) Department of Physics, Tamkang University, Tamsui Dist., New Taipei 25137, Taiwan, ROC\\ \({^3}\) Hon Hai Research Institute, Taipei, Taiwan, ROC\\ \({^4}\) Physics Division, National Center for Theoretical Sciences, Taipei, Taiwan, ROC}

\begin{abstract}

We investigate the effectiveness of the stabilizer R\'enyi entropy (SRE), a quantifier associated with non-stabilizer resources (quantum magic), as an indicator of quantum phase transitions. Specifically, we analyze the behavior of the purity-corrected SRE of reduced density matrices in the ground states of two one-dimensional spin models: the transverse axial next-nearest-neighbor Ising (TANNNI) model and the quantum compass model (QCM). The ground state of the TANNNI model is obtained using exact diagonalization techniques, while the QCM is analyzed using the Jordan--Wigner (JW) transformation followed by Bogoliubov diagonalization of the resulting quadratic fermionic Hamiltonian. For the TANNNI model, the purity-corrected SRE successfully detects the antiphase--floating phase transition in the high-frustration regime, while in the low-frustration regime the raw (purity-uncorrected) SRE reproduces the known ferromagnetic--paramagnetic phase boundaries more accurately. For the QCM, the purity-corrected SRE exhibits a clear signature near the isotropic point \(J_x/J_z=1\), where the system undergoes a first-order quantum phase transition. Our results establish SRE of reduced states as a complementary probe of quantum criticality and provide further insight into the role of non-stabilizer resources in many-body quantum phase transitions.

 \vspace*{.3cm}
 \noindent \(^\dagger\){These authors contributed equally to this work.}
 \vspace*{.3cm}\\
 \noindent \(^\star\)Correspondence could be addressed to \href{mailto:georgebsws@gmail.com}{georgebsws@gmail.com} and \href{mailto:anindya@nitsikkim.ac.in}{anindya@nitsikkim.ac.in}.

\end{abstract}
\maketitle

\section{Introduction}

Quantum phase transitions (QPTs) arise in many-body quantum systems when a qualitative change in the ground-state properties occurs due to the variation of an external control parameter or coupling strength at absolute zero temperature~\cite{sachdev_2011}. Unlike classical phase transitions driven by thermal fluctuations, QPTs are governed by quantum fluctuations and the restructuring of quantum correlations in the ground state. Over the past decades, paradigmatic quantum spin models have served as fundamental platforms for investigating the interplay between interactions, correlations, and QPTs~\cite{sachdev_2011,Vojta_2003,Dutta_Sen_2015}.

With the rapid development of quantum information theory, various quantum resource measures have emerged as powerful tools to characterize many-body quantum systems. Quantum entanglement, in particular, has been extensively studied as an indicator of quantum criticality, where measures such as concurrence, entanglement entropy, geometric measure of entanglement, {\ab and others} exhibit distinctive signatures near critical points in several quantum spin models~\cite{PhysRevA.66.032110,Osterloh2002,Gu_2006,PhysRevLett.90.227902,PhysRevA.70.042311,PhysRevLett.93.250404,dutta_aeppli_chakrabarti_divakaran_rosenbaum_sen_2015,PhysRevA.90.032301,Biswas_2020,Mondal_2023}. Beyond entanglement, other informational quantities, including quantum discord, quantum coherence, shared purity, Bell nonlocality, and quantum Fisher information, have also been employed to detect and characterize QPTs, further demonstrating that quantum phase transitions are intrinsically connected to the redistribution of quantum resources in many-body systems~\cite{PhysRevB.78.224413,PhysRevA.80.022108,Li2016,PhysRevB.101.115142,Du2022,Biswas_2024,Huang2013,PhysRevA.106.022208,PhysRevA.90.062129,YE2016151,doi:10.1142/S0219749920500161,Tan2025}.

More recently, quantum magic has gained considerable attention as a fundamental resource enabling universal fault-tolerant quantum computation within the stabilizer formalism~\cite{Veitch_2014,PhysRevLett.118.090501}. In the stabilizer framework, quantum circuits composed solely of Clifford gates acting on stabilizer states can be efficiently simulated using classical computers, as formalized by the Gottesman--Knill theorem~\cite{gottesman1997stabilizercodesquantumerror,gottesman1998heisenbergrepresentationquantumcomputers,PhysRevA.70.052328,PhysRevLett.116.250501,Bergou2021,Biswas_quanta}. The presence of non-stabilizer resources, commonly referred to as magic, is essential to achieve universal quantum computation. Several quantifiers of magic have been introduced, including mana, robustness of magic, and relative entropy of magic; however, these measures often require computationally demanding optimizations~\cite{PhysRevA.97.062332,Heinrich2019robustnessofmagic,10.1098/rspa.2019.0251,PhysRevLett.124.090505,PRXQuantum.2.010345,PRXQuantum.3.020333,Dai2022}. Stabilizer R\'enyi Entropy (SRE) provides a computationally accessible and physically meaningful measure to quantify non-stabilizer resources in quantum states~\cite{PhysRevLett.128.050402,Haug2023stabilizerentropies,PhysRevA.110.L040403}.

Entanglement and quantum magic represent fundamentally distinct nonclassical resources that arise in different operational and computational settings. {\ab Extensive studies have demonstrated that entanglement measures serve as sensitive probes of quantum phase transitions in a variety of spin models~\cite{PhysRevA.66.032110,Osterloh2002,Gu_2006,PhysRevLett.90.227902,PhysRevA.70.042311,PhysRevLett.93.250404,dutta_aeppli_chakrabarti_divakaran_rosenbaum_sen_2015,PhysRevA.90.032301,Biswas_2020,Mondal_2023}.} Entanglement has been recognized as a central quantity in the characterization of quantum many-body systems and plays a crucial role in determining the classical simulability of quantum states within tensor-network-based frameworks such as matrix product state (MPS) simulations. The efficiency of MPS methods is strongly constrained by the growth of bipartite entanglement~\cite{PhysRevLett.91.147902,eisert2013entanglementtensornetworkstates}. 
{\ab On the other hand,} quantum magic quantifies the non-stabilizer resource content of quantum states and restricts the classical simulability of universal quantum computation within the stabilizer framework. States lacking magic can be efficiently simulated using stabilizer techniques, whereas the presence of magic introduces additional computational overhead~\cite{Bravyi2019simulationofquantum,deSilva_2024}. Although entanglement and magic originate from different theoretical frameworks, both resources impose fundamental restrictions on the classical simulation of universal quantum computation.
Recent investigation in discrete-time quantum walks have reported scenarios in which high entanglement coincides with reduced magic, suggesting that these resources capture distinct aspects of quantum correlations~\cite{7rwg-lhpv}. These observations motivate the exploration of SRE as an independent probe for many-body quantum phenomena.

In this work, we investigate SRE as an indicator of quantum phase transitions. Specifically, we analyze SRE evaluated on reduced density matrices obtained from the ground states of the TANNNI model and the QCM. While entanglement-based probes have been studied in these models, our focus here is to examine whether magic-based probes can capture signatures of critical behavior. We adopt reduced density matrix approach to facilitate efficient computation while preserving essential local correlation properties of the many-body ground state.

We quantify the Stabilizer R\'enyi Entropy (SRE), which measures the distance of a quantum state from the stabilizer polytope. For an $n$-qubit density matrix $\rho$, the second-order Stabilizer R\'enyi Entropy is defined as
\begin{equation} \label{e2}
M_2(\rho)=-\log_2\left(\frac{1}{2^n}\sum_{P\in \mathcal{P}_n}\left[\mathrm{Tr}(\rho P)\right]^4\right),
\end{equation}
where $\mathcal{P}_n$ denotes the $n$-qubit Pauli group consisting of all tensor products of single-qubit Pauli operators, including identity. The quantity $\mathrm{Tr}(\rho P)$ corresponds to the expectation value of the Pauli operator $P$ in the state $\rho$. {For pure states,} SRE vanishes for stabilizer states and takes positive values for non-stabilizer states, thereby providing a quantitative measure of quantum magic~\cite{PhysRevLett.128.050402,Haug2023stabilizerentropies,PhysRevA.110.L040403}.

{Note that, our motivation is to study magic based probe of reduced density matrices of one-{\ab qubit} or two-qubits after tracing out rest of the system. These reduced density matrices are not pure states and for mixed reduced density matrices, SRE is not a magic monotone. For mixed states \(M_2\) can be viewed as a measure of {\ab non-flatness} {\ab of} the Pauli expectation values~\cite{10.21468/SciPostPhys.19.4.085}. Magic is related to non-flatness of Pauli expectation values. However, \(M_2\) {\ab for} mixed state{\ab s} contains contributions both from non-stabilizerness and mixedness~\cite{10.21468/SciPostPhys.19.4.085}. Therefore, for the mixed state case the purity corrected second-order stabilizer R\'enyi entropy {\ab is used. It is} defined {\ab as}~\cite{PhysRevLett.128.050402,10.21468/SciPostPhys.19.4.085}
\begin{equation}
\widetilde M_2(\rho)=M_2(\rho)-S_2(\rho),
\end{equation}
where \(S_2(\rho)=-\log_2\Tr(\rho^2)\) is the second{\ab -order} R\'enyi entropy. Equivalently,
\begin{equation} \label{e3}
\widetilde M_2(\rho)
=
-\log_2\left(
\frac{1}{2^n\,\Tr(\rho^2)}
\sum_{P\in\mathcal P_n}\left[\Tr(\rho P)\right]^4
\right).
\end{equation}
\(\widetilde{M_2}\) tells how non-stabilizer-like the reduced density matrix is relative to states with same purity. Note that when we partially trace out a subsystem, mixed-ness in the {\ab reduced} subsystem comes from the entanglement between the two subsystems. 
This purity corrected SRE is an efficiently computable probe of magic in the subsystems of many-body {\ab systems}, however {\ab it is} not a magic monotone for mixed state{\ab s}~\cite{10.21468/SciPostPhys.19.4.085}. For a genuine mixed-state magic monotone, Leone and Bittel instead propose a convex-roof extension of stabilizer entropies~\cite{PhysRevA.110.L040403}. {\ab We choose to study the} purity corrected second-order stabilizer R\'enyi entropy \(\widetilde{M_2}\) {\ab in the aforementioned systems}.}

From now on, we drop the term `second-order' as we only consider the second-order stabilizer R\'enyi entropy {{and purity corrected second-order stabilizer R\'enyi entropy}}.

Note that the summation in Eq.~(\ref{e2}) and Eq.~(\ref{e3}) runs over all $4^n$ elements of the $n$-qubit Pauli group, which leads to exponential computational complexity as the system size increases. This exponential scaling makes the direct evaluation of SRE for large many-body systems numerically challenging. Recently, an exact algorithm capable of computing SRE for systems up to approximately $20$ qubits was introduced in Ref.~\cite{huang2026fastexactapproachstabilizer}. In addition, Ref.~\cite{PhysRevLett.131.180401} proposed a Pauli sampling approach that enables efficient estimation of SRE within the MPS framework. In that work, the authors computed the SRE for the transverse field Ising model with up to $14$ qubits and validated their numerical results against analytical solutions obtained via mapping the model to free fermions~\cite{PhysRevLett.131.180401}.
In contrast to these approaches, our reduced density matrix approach significantly simplifies the computational effort, as the Pauli group dimension is {{just $4$ for one-qubit reduced density matrix or \(4^2 = 16\) for two-qubit reduced density matrix,}} thereby avoiding the exponential scaling associated with the full many-body density matrix. Previous studies have demonstrated that magic-based resource measures can detect quantum criticality in spin models such as the XY, TFI, and XXZ chains~\cite{PhysRevA.106.062405,pyzr-jmvw,He2023}. More recently, stabilizer R\'enyi entropy has also been investigated in strongly correlated fermionic systems in the thermodynamic limit. In Ref.~\cite{Zhang2026}, a large-\(N\) framework for computing the SRE in coupled Sachdev--Ye--Kitaev models was developed, and a series of first-order SRE transitions were identified, including transitions that are not visible in conventional thermodynamic quantities. These results further support the role of SRE as an independent probe of many-body quantum phenomena. Motivated by these results, we focus on the transverse axial next-nearest-neighbor Ising model and the quantum compass model, both of which exhibit multiple quantum phases separated by critical transitions.


In the upcoming sections~\ref{II} and~\ref{III} we analyze the SRE (purity corrected/uncorrected) of ground states for the TANNNI model and the QCM, respectively. We present a conclusion in Sec.~\ref{IV}.

\section{Transverse Axial Next-Nearest-Neighbor Ising Model}\label{II}

The Hamiltonian of the TANNNI model can be written as 
\begin{equation}
H_{\mathrm{TANNNI}} = -J_1 \sum_{i=1}^{N} \sigma_i^{z}\sigma_{i+1}^{z} 
+ J_2 \sum_{i=1}^{N} \sigma_i^{z}\sigma_{i+2}^{z} 
- \gamma \sum_{i=1}^{N} \sigma_i^{x},
\end{equation}
where $J_1$ and $J_2$ denote the nearest-neighbor and next-nearest-neighbor coupling strengths, respectively, and $\gamma$ is the transverse magnetic field. 

\begin{figure}
\centering
\begin{tikzpicture}[scale=1.2]

\draw[->] (0,0) -- (5.25,0) node[right] {$J_2/J_1$};
\draw[->] (0,0) -- (0,5.25) node[above] {$\gamma/J_1$};

\node at (-0.2,-0.1) {0};
\node at (2.5,-0.3) {0.5};
\node at (-0.4,5) {1.0};

\draw[thick] (0,5) -- (2.5,0);          
\draw[thick] (2.5,0) -- (5,2.5);   

\draw[thick, smooth]
plot coordinates {
(1.5,2.0)
(1.78,1.6)
(2.05,1.3)
(2.25,1.2)
(2.6,1.3)
(3.25,1.85)
(4.1,2.5)
};

\node at (1.1,1.0) {Ferromagnetic};
\node at (4.2,1.0) {Antiphase};
\node[rotate=45] at (3.4,1.5) {Floating};
\node at (2.5,2.5) {Paramagnetic};

\node at (1.08,0.6) {$++++++++$};
\node at (4.2,0.6) {$++--++--$};

\node at (2.0,0.5) {};

\node at (3.9,0.5) {};

\end{tikzpicture}
\caption{Phase diagram of the TANNNI model according to~\cite{PhysRevE.75.021105}.}
\label{schematic}
\end{figure}
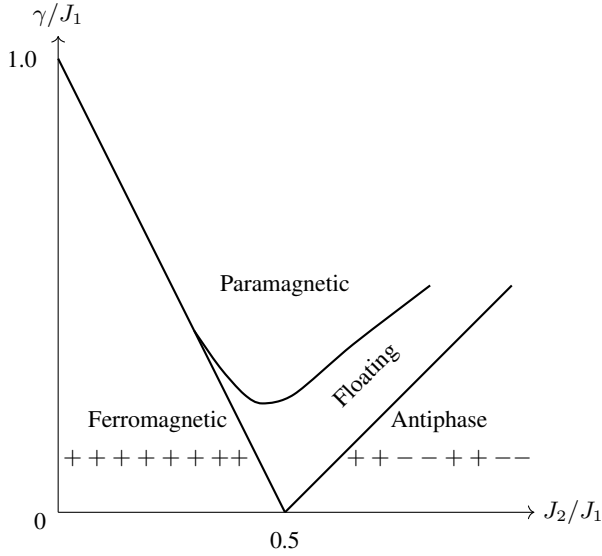

A series of studies conducted over the past three decades have progressively unveiled the rich quantum phase diagram of the TANNNI model~\cite{Arizmendi1991,Sen1995,1996ZPhyB.101..597R,Rieger1996,PhysRevB.67.094435,PhysRevE.75.021105,Nagy_2011}. The competition between nearest-neighbor ferromagnetic interactions and next-nearest-neighbor antiferromagnetic interactions introduces frustration, giving rise to multiple quantum phases and intricate phase boundaries. 
A representation of the phase diagram is shown in Fig.~\ref{schematic}, which summarizes the qualitative structure of the phase boundaries as a function of the transverse field strength and frustration parameter.

\begin{figure}
    \centering
       \includegraphics[width=1\linewidth]{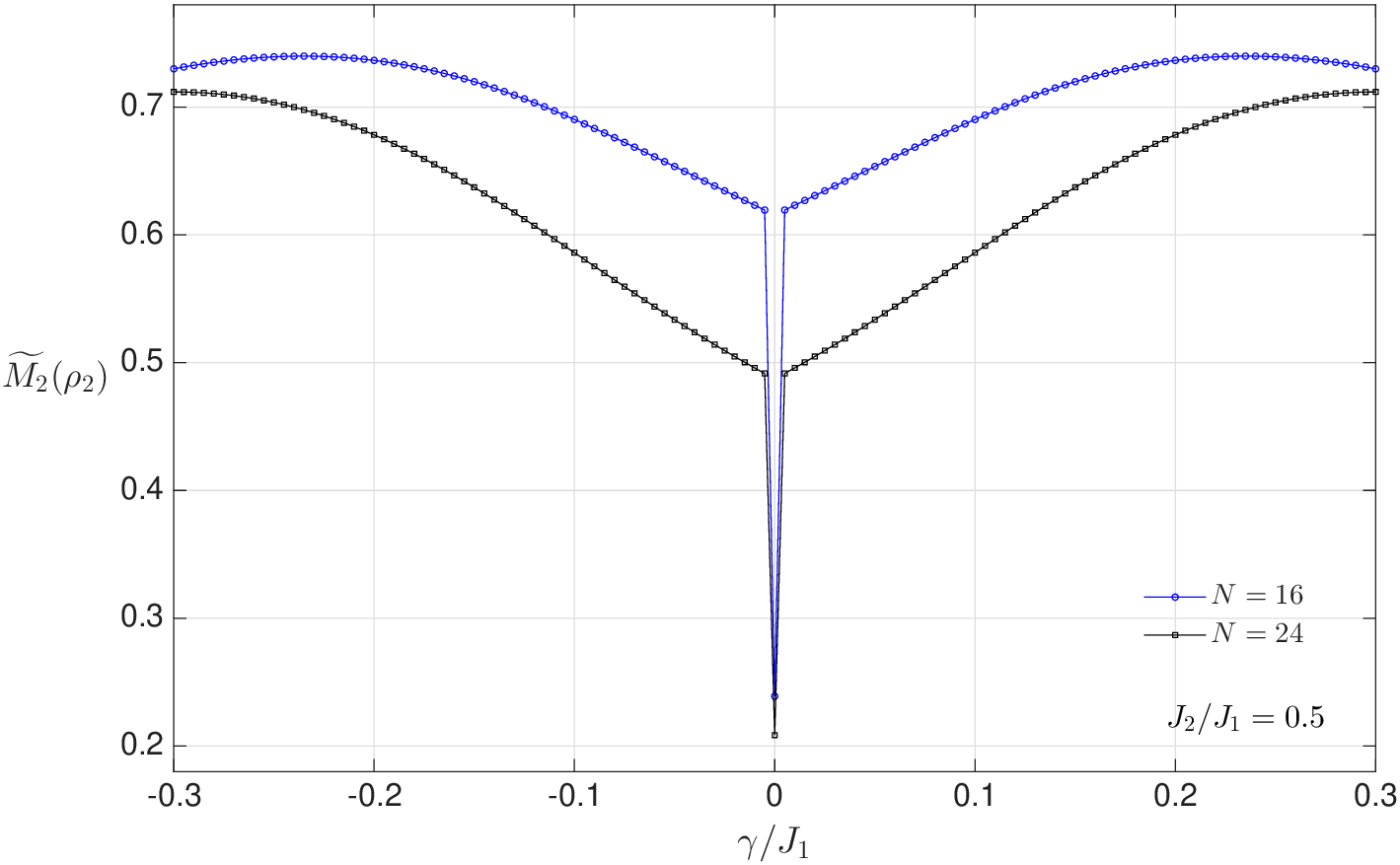}
    \caption{Purity-corrected stabilizer R\'enyi entropy of the two-site reduced density matrix $\widetilde{M_2}\left(\rho_2\right)$ plotted as a function of the transverse field $\gamma/J_1$ for fixed $J_2/J_1 = 0.5$.}
    \label{Multicriticality_2}
\end{figure}

We first investigate whether the purity corrected SRE $\widetilde{M_2}$ of two-site reduced density matrix display{\ab s} any indication for the presence of the multi-critical point at \(J_2/J_1=0.5\) and \(\gamma/J_1=0\). We plot purity corrected SRE (\(\widetilde{M_2}\)) varying the transverse field \(\gamma/J_1\) keeping \(J_2/J_1=0.5\) in Fig.~\ref{Multicriticality_2}.
We observe that the variation of $\widetilde{M_2}$ is mirror symmetric about the critical point $\gamma/J_1=0$ and {\ab displays} a dip at the same critical point. Therefore, the variation of $\widetilde{M_2}$ {\ab signals} the underlying phase transition.

\begin{figure}
    \centering
       \includegraphics[width=0.99\linewidth]{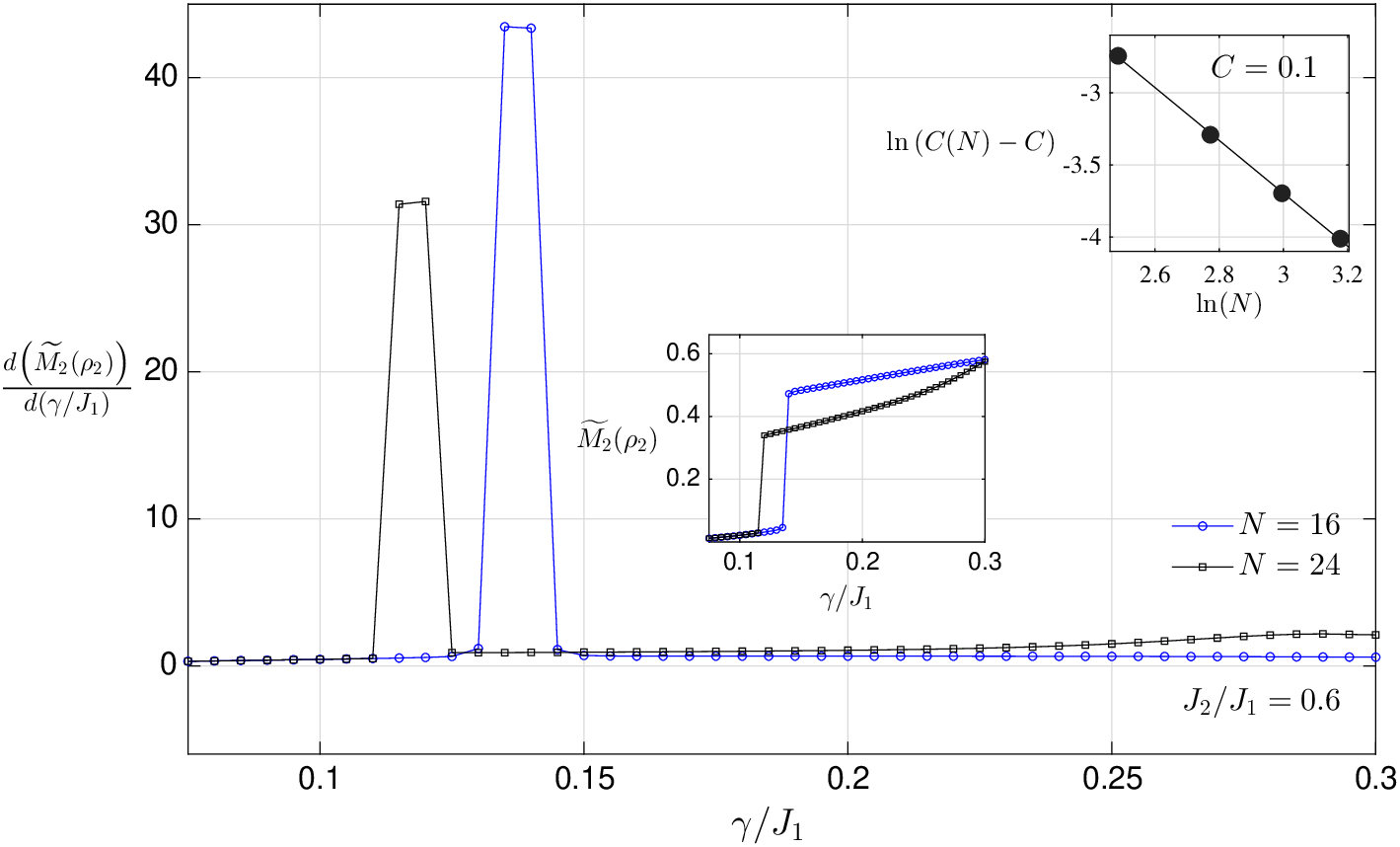}
   (a) \includegraphics[width=1\linewidth]{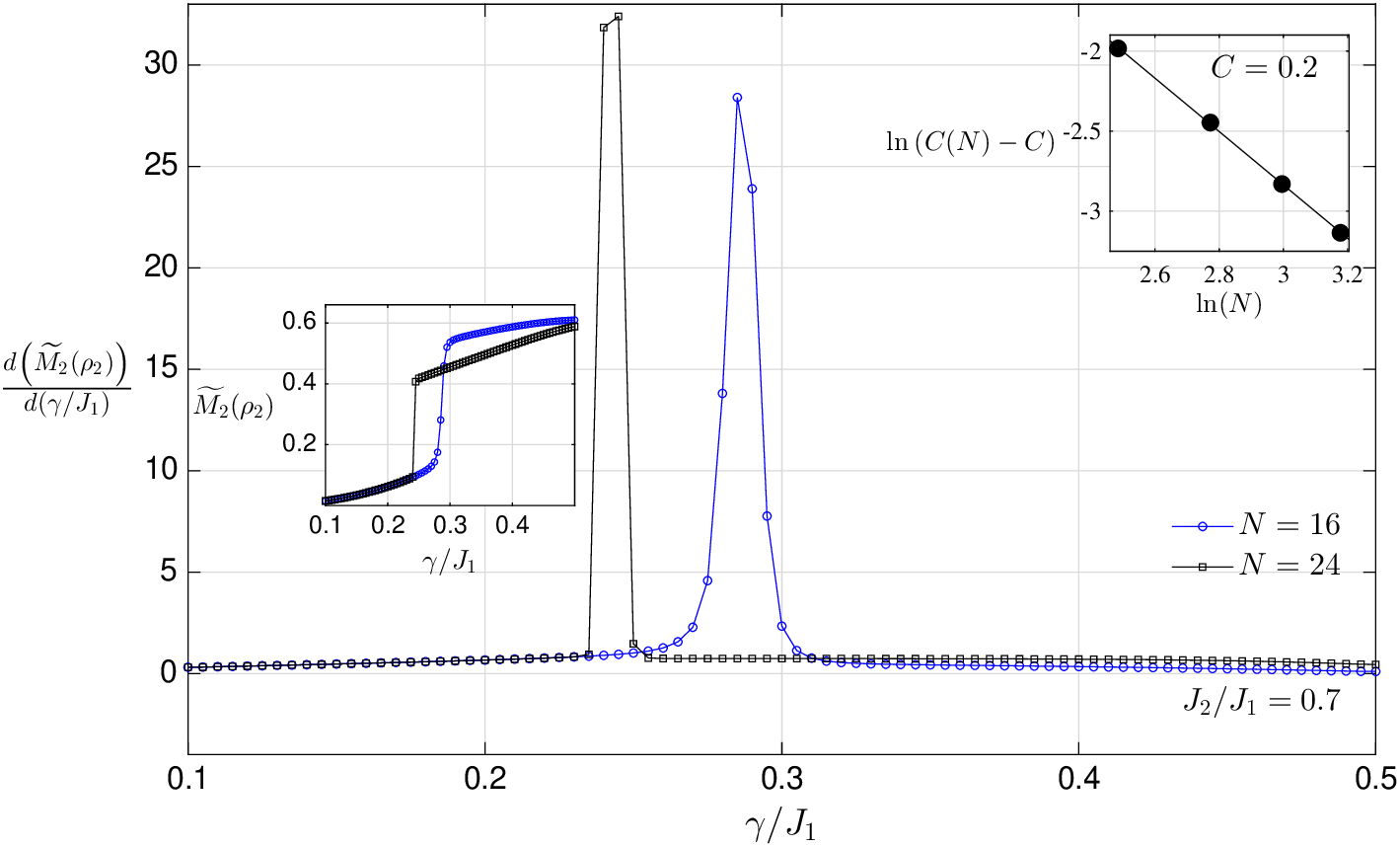}
   (b) 
    \caption{Purity corrected stabilizer Rényi entropy (in left inset) {{of two qubit reduced density matrix}} ($\widetilde{M_2}(\rho_2)$) and its first derivative with respect to the ratio of the transverse field strength to the nearest neighbor interaction strength, $\frac{d\left(\widetilde{M_2}(\rho_2)\right)}{d(\gamma/J_1)}$, plotted as functions of $\gamma/J_1$ for different frustration parameters: (a) $J_2/J_1 = 0.6$, (b) $J_2/J_1 = 0.7$. Each panel shows results for two system sizes, $N = 16$ (blue curves) and $N = 24$ (black curves). The right inset panels display the finite-size scaling behavior of the maxima in $\frac{d\left(\widetilde{M_2}(\rho_2)\right)}{d(\gamma/J_1)}$, demonstrating the convergence of the finite-size critical points $C(N)$ toward the known thermodynamic-limit values with increasing system size. The known quantum critical points ($C$) corresponding to each frustration parameter are indicated in the respective right inset panels.
}
    \label{right_ANNNI_sre-re_1}
\end{figure}

\begin{figure}
    \centering
       \includegraphics[width=0.99\linewidth]{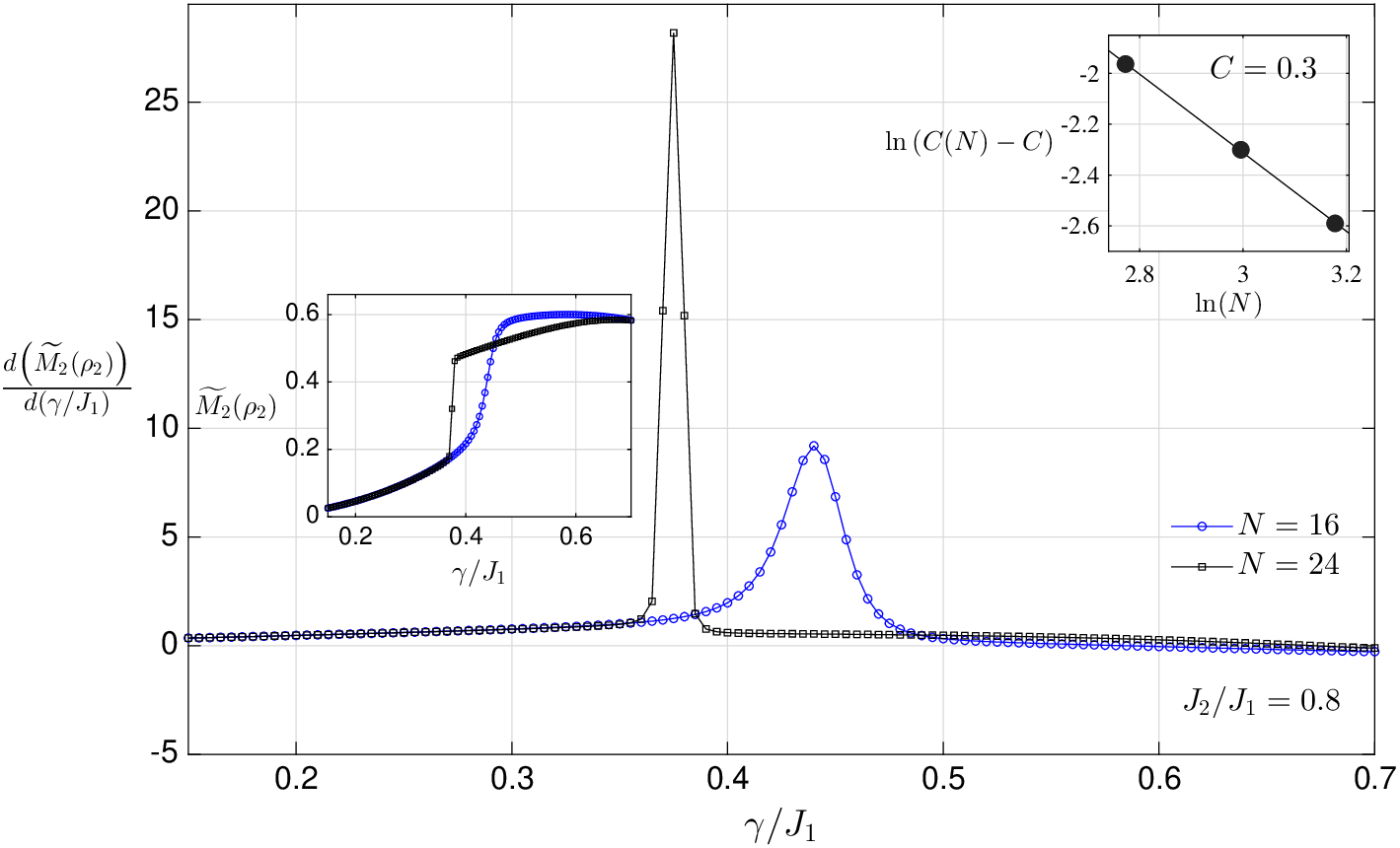}
   (a) \includegraphics[width=1\linewidth]{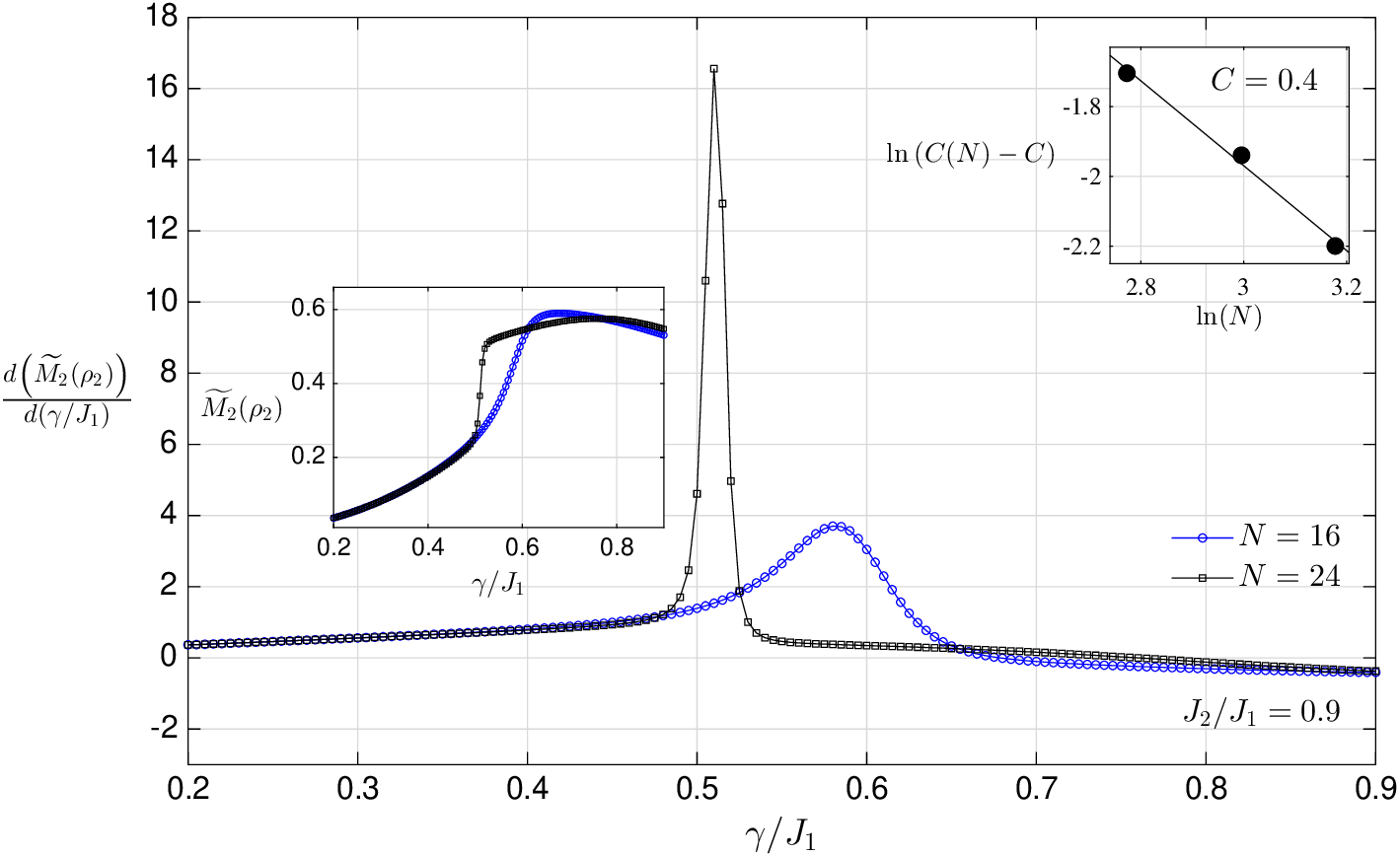}
   (b) 
    \caption{Purity corrected stabilizer Rényi entropy (in left inset) {{of two qubit reduced density matrix}} ($\widetilde{M_2}(\rho_2)$) and its first derivative with respect to the ratio of the transverse field strength to the nearest neighbor interaction strength, $\frac{d\left(\widetilde{M_2}(\rho_2)\right)}{d(\gamma/J_1)}$, plotted as functions of $\gamma/J_1$ for different frustration parameters: (a) $J_2/J_1 = 0.8$, (b) $J_2/J_1 = 0.9$. Each panel shows results for two system sizes, $N = 16$ (blue curves) and $N = 24$ (black curves). The right inset panels display the finite-size scaling behavior of the maxima in $\frac{d\left(\widetilde{M_2}(\rho_2)\right)}{d(\gamma/J_1)}$, demonstrating the convergence of the finite-size critical points $C(N)$ toward the known thermodynamic-limit values with increasing system size. The known quantum critical points ($C$) corresponding to each frustration parameter are indicated in the respective right inset panels
}
    \label{right_ANNNI_sre-re_2}
\end{figure}

{\ab In} the higher frustration region \textit{i.e.}, for \(J_2/J_1>0.5\) we perform {\ab a} similar analysis and investigate whether the behavior of the purity corrected SRE can indicate the phase transition from the antiphase to the floating phase. In Fig.~\ref{right_ANNNI_sre-re_1}, we plot $\widetilde{M_2}$ and its derivative as functions of the transverse field strength \(\gamma/J_1\) for $J_2/J_1 = 0.6, 0.7$ and we plot the same for $J_2/J_1 = , 0.8, 0.9$ in Fig.~\ref{right_ANNNI_sre-re_2}. The corresponding critical points in the $\gamma/J_1$ axis in this region are $C=0.1, 0.2, 0.3,$ and $0.4$, respectively.

Our numerical results reveal that $\widetilde{M_2}$ exhibits a noticeable change in curvature in the vicinity of the critical points. Correspondingly, the first derivative of $\widetilde{M_2}$ develops a pronounced maxima near $C$. Importantly, as the number of spins in the chain ($N$) increases, these maxima systematically approach the known critical points, indicating that $\widetilde{M_2}$ captures the underlying critical behavior. To visualize this observation, we perform a finite-size scaling analysis and the results are displayed in the right inset of each panel in Fig.~\ref{right_ANNNI_sre-re_1} and in Fig.~\ref{right_ANNNI_sre-re_2}. 
In the finite-size scaling analysis, we fit a linear relation between $\ln(C(N) - C)$ and $\ln N$, where $C(N)$ are the position of the peak in the first derivative of $\widetilde{M_2}$ for a particular $N$. 
A good linear fit indicates reliable estimation of the critical point.

\begin{figure}
    \centering
       \includegraphics[width=0.99\linewidth]{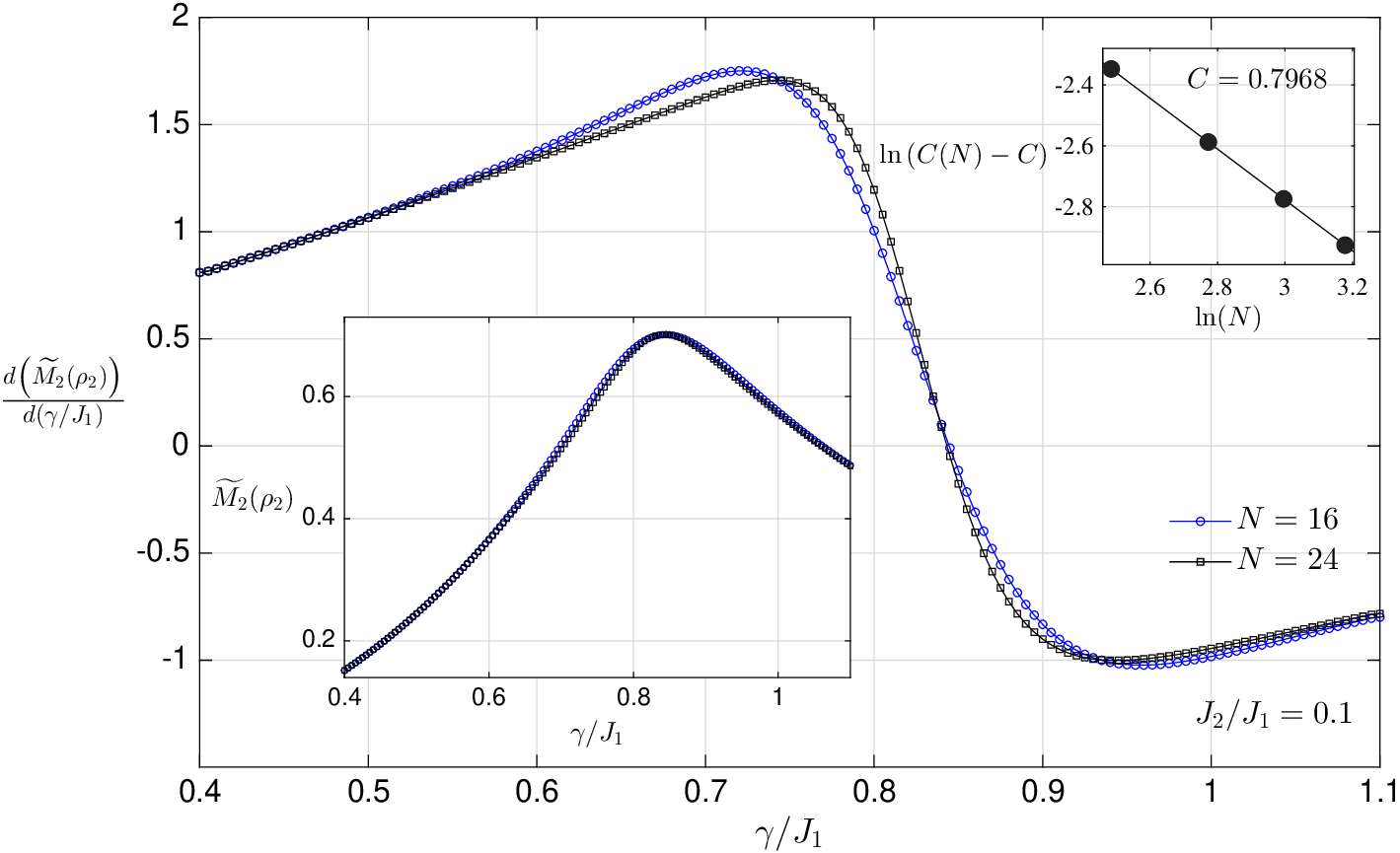}
   (a) \includegraphics[width=1\linewidth]{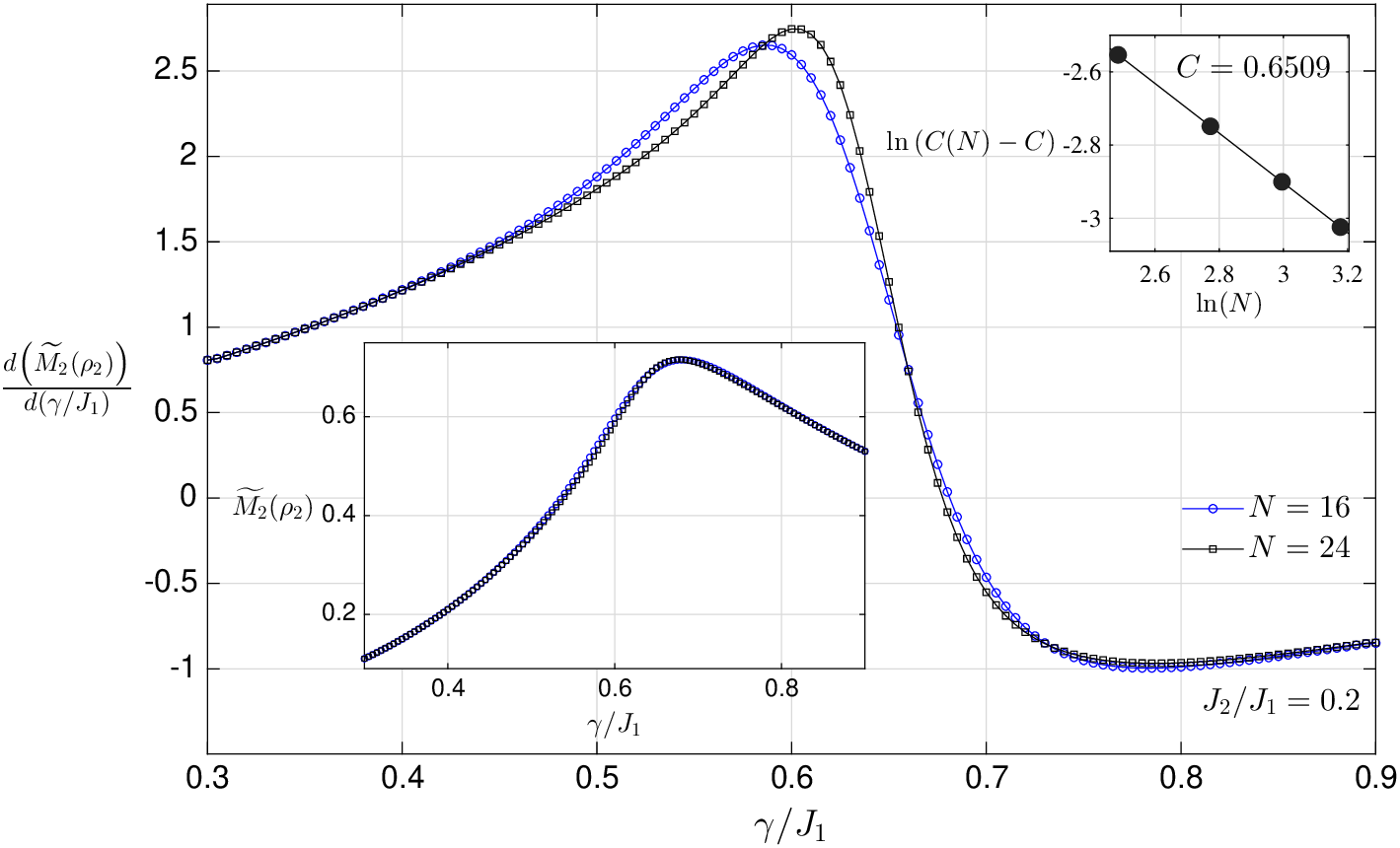}
   (b) 
    \caption{Purity corrected stabilizer Rényi entropy (in left inset) {{of two qubit reduced density matrix}} ($\widetilde{M_2}(\rho_2)$) and its first derivative with respect to the ratio of the transverse field strength to the nearest neighbor interaction strength, $\frac{d\left(\widetilde{M_2}(\rho_2)\right)}{d(\gamma/J_1)}$, plotted as functions of $\gamma/J_1$ for different frustration parameters: (a) $J_2/J_1 = 0.1$, (b) $J_2/J_1 = 0.2$. Each panel shows results for two system sizes, $N = 16$ (blue curves) and $N = 24$ (black curves). The right inset panels display the finite-size scaling behavior of the maxima in $\frac{d\left(\widetilde{M_2}(\rho_2)\right)}{d(\gamma/J_1)}$. The critical points in the thermodynamic-limit are estimated from the finite-size scaling behavior. The estimated thermodynamic-limit critical points ($C$) corresponding to each frustration parameter are indicated in the respective right inset panels.
}
    \label{left_ANNNI_sre-re_1}
\end{figure}

\begin{figure}
    \centering
       \includegraphics[width=0.99\linewidth]{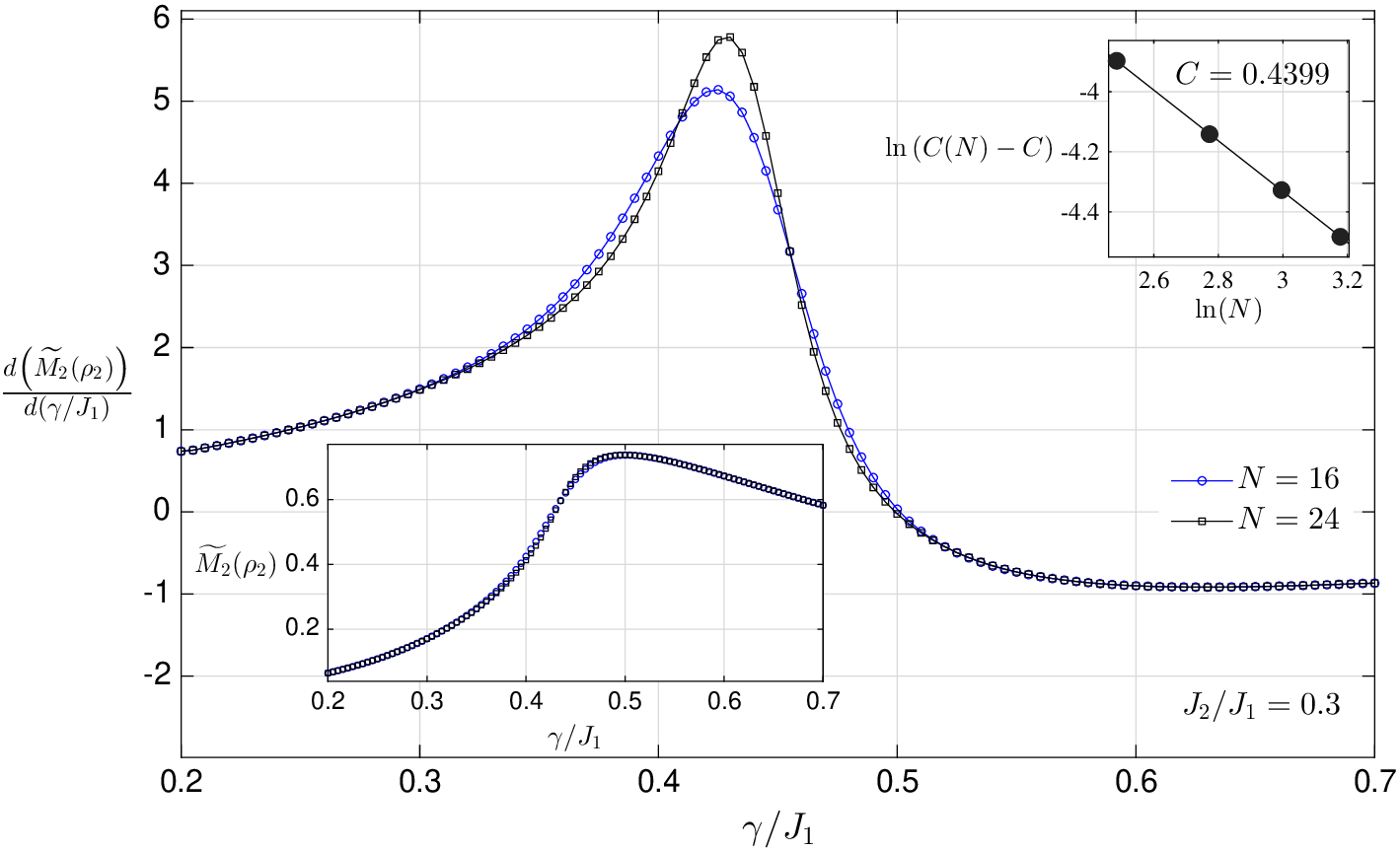}
   (a) \includegraphics[width=1\linewidth]{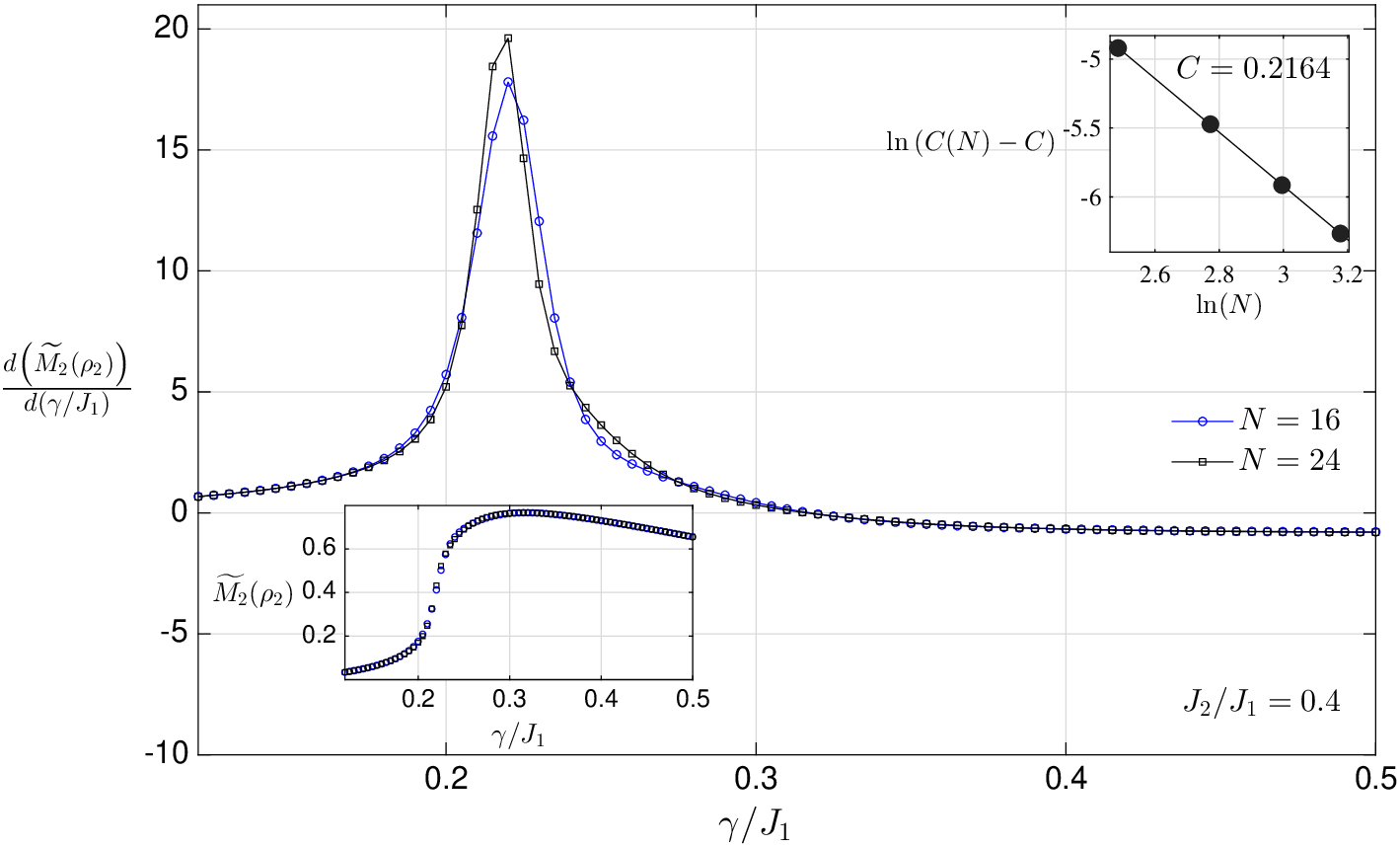}
   (b) 
    \caption{Purity corrected stabilizer Rényi entropy (in left inset) {{of two qubit reduced density matrix}} ($\widetilde{M_2}(\rho_2)$) and its first derivative with respect to the ratio of the transverse field strength to the nearest neighbor interaction strength, $\frac{d\left(\widetilde{M_2}(\rho_2)\right)}{d(\gamma/J_1)}$, plotted as functions of $\gamma/J_1$ for different frustration parameters: (a) $J_2/J_1 = 0.3$, (b) $J_2/J_1 = 0.4$. Each panel shows results for two system sizes, $N = 16$ (blue curves) and $N = 24$ (black curves). The right inset panels display the finite-size scaling behavior of the maxima in $\frac{d\left(\widetilde{M_2}(\rho_2)\right)}{d(\gamma/J_1)}$. The critical points in the thermodynamic-limit are estimated from the finite-size scaling behavior. The estimated thermodynamic-limit critical points ($C$) corresponding to each frustration parameter are indicated in the respective right inset panels.
}
    \label{left_ANNNI_sre-re_2}
\end{figure}

We then study the crital transitions of the purity corrected SRE of the two-qubit reduced density matrix in the low frustration region, \(0<J_2/J_1<0.5\). In Fig.~\ref{left_ANNNI_sre-re_1} and in Fig.~\ref{left_ANNNI_sre-re_2} we present the behavior of $\widetilde{M_2}$ and its first derivative with respect to the ratio of the transverse field strength and the nearest neighbor interaction strength $\gamma/J_1$ for different values of the frustration parameter, namely $J_2/J_1 = 0.1, 0.2, 0.3,$ and $0.4$. Although we observe noticeable change in curvature, finite size scaling analysis reveals that they {\ab do not converge towards} the known critical points. The critical points with respect to the transition of $\widetilde{M_2}$ in the $\gamma/J_1$ axis for $J_2/J_1=0.1, 0.2, 0.3, \text{and}$ 0.4 are found to be $C=0.7968, 0.6509, 0.4399,$ and $0.2164$, respectively. The scaling behavior is displayed in the right inset of each panel of Fig.~\ref{left_ANNNI_sre-re_1} and Fig.~\ref{left_ANNNI_sre-re_2}.

\begin{figure}
\centering
\begin{tikzpicture}[scale=1.2]

\draw[->] (0,0) -- (5.25,0) node[right] {$J_2/J_1$};
\draw[->] (0,0) -- (0,5.25) node[above] {$\gamma/J_1$};

\node at (-0.2,-0.1) {0};
\node at (2.5,-0.3) {0.5};
\node at (-0.4,5) {1.0};

\draw[thick] (0,5) -- (2.5,0);
\draw[thick] (2.5,0) -- (5,2.5);

\draw[thick, smooth]
plot coordinates {
(1.5,2.0)
(1.78,1.6)
(2.05,1.3)
(2.25,1.2)
(2.6,1.3)
(3.25,1.85)
(4.1,2.5)
};

\draw[ smooth]
plot coordinates {
(0.0/2,4.22)
(0.25/2,4.19)
(0.50/2,4.1385)
(0.75/2,4.0745)
(1.00/2,3.984)
(1.25/2,3.853)
(1.50/2,3.695)
(1.75/2,3.4765)
(2.00/2,3.2545)
(2.25/2,2.9955)
(2.50/2,2.727)
(2.75/2,2.460)
(3.00/2,2.1995)
(3.25/2,1.926)
(3.50/2,1.639)
(3.75/2,1.361)
(4.00/2,1.082)
(4.25/2,0.801)
(4.50/2,0.524)
(4.75/2,0.256)
(5.00/2,0.0)
};

\draw[
    black,
    only marks,
    mark=*,
    mark size=0.5pt
]
plot coordinates {
(0.25/2,4.205)
(0.50/2,4.1385)
(0.75/2,4.0745)
(1.00/2,3.984)
(1.25/2,3.853)
(1.50/2,3.695)
(1.75/2,3.4765)
(2.00/2,3.2545)
(2.25/2,2.9955)
(2.50/2,2.727)
(2.75/2,2.460)
(3.00/2,2.1995)
(3.25/2,1.926)
(3.50/2,1.639)
(3.75/2,1.361)
(4.00/2,1.082)
(4.25/2,0.801)
(4.50/2,0.524)
(4.75/2,0.256)
(5.00/2,0.0)
(6.00/2,1/2)
(7.00/2,2/2)
(8.00/2,3/2)
(9.00/2,4/2)
};

\draw[
    blue,
    only marks,
    mark=triangle,
    mark size=1.9pt
]
plot coordinates {
(0.00/2,5)
(1.00/2,4)
(2.00/2,3)
(3.00/2,2)
(4.00/2,1)
(5.00/2,0.0)
(6.00/2,1/2)
(7.00/2,2/2)
(8.00/2,3/2)
(9.00/2,4/2)
};


\draw[->, thick]
(3.8/2,3.5) -- (2.55/2,2.75);

\draw[
    black,
    only marks,
    mark=*,
    mark size=0.5pt
]
plot coordinates {
(3.1/2,3.75)};

\draw[ smooth]
plot coordinates {
(2.9/2,3.75)
(3.3/2,3.75)};

\node at (4.7/2,3.75) {$\left(\widetilde{M_2}\right)_{\text{transition}}$};

\draw[->, thick]
(3.6/2,4.65) -- (1.11/2,4);
\draw[->, thick]
(3.6/2,4.65) -- (2.07/2,3.04);

\draw[
    blue,
    only marks,
    mark=triangle,
    mark size=1.9pt
]
plot coordinates {
(2.9/2,4.8)};

\node at (4.4/2,4.8) {$\left({\color{black}{M_2}}\right)_{\text{transition}}$};

\node at (1.1,1.0) {Ferromagnetic};
\node at (4.2,1.0) {Antiphase};
\node[rotate=45] at (3.4,1.5) {Floating};
\node at (2.5,2.5) {Paramagnetic};

\node at (1.08,0.6) {$++++++++$};
\node at (4.2,0.6) {$++--++--$};

\end{tikzpicture}
\caption{Phase diagram of the TANNNI model from~\cite{PhysRevE.75.021105}, together with the transition lines extracted from the reduced-state stabilizer R\'enyi entropy. The thin black curve with circular markers represents the transition points obtained from the purity-corrected SRE \(\widetilde{M_2}\), while the blue triangle markers denote the transition points extracted from the raw (purity-uncorrected) SRE \(M_2\). The \(\widetilde{M_2}\)-based transition shows a mismatch with the established phase boundaries for \(J_2/J_1 < 0.5\), whereas for \(J_2/J_1 \geq 0.5\) it aligns well with the antiphase--floating phase boundary. In contrast, the transition points obtained from the raw \(M_2\) are found to agree with the known quantum phase boundaries in both the low- and high-frustration regimes.
}
\label{schematic_2}
\end{figure}
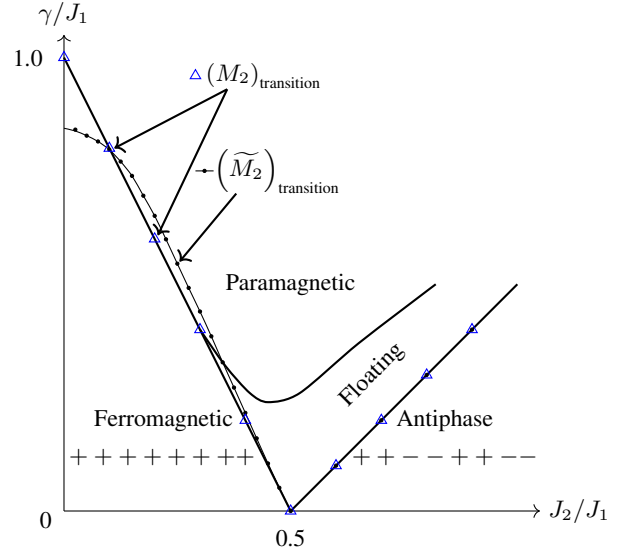

We then perform a detailed finite-size scaling analysis to determine the critical transition points obtained from the two-qubit \(\widetilde{M_2}\) for \(J_2/J_1 = 0.025, 0.050, \cdots, 0.475\). The resulting transition line is compared with the known quantum phase boundary in Fig.~\ref{schematic_2}. From this comparison, we observe that in the region \(0.4 < J_2/J_1 < 0.5\), the critical transition points extracted from \(\widetilde{M_2}\) lie very close to the established quantum phase transition points. In the intermediate regime \(0.1 < J_2/J_1 < 0.4\), the \(\widetilde{M_2}\) transition points show noticeable deviations from the known phase boundary, while in the {\ab very} low-frustration regime \(0 < J_2/J_1 < 0.1\), the extracted transition points differ significantly from the established quantum critical line.

\begin{figure}
     \centering
     \includegraphics[width=0.99\linewidth]{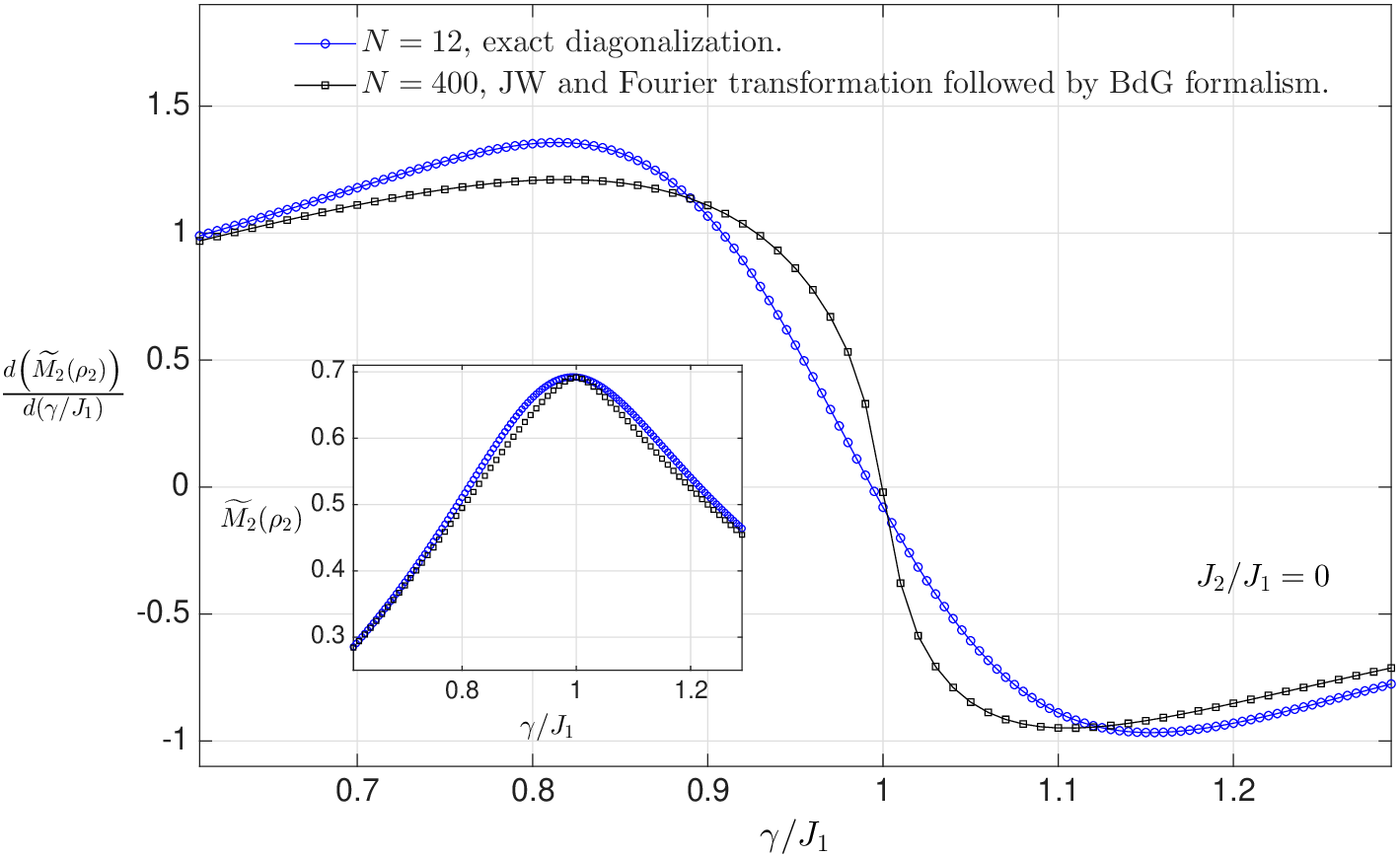}
     \caption{Purity corrected stabilizer Rényi entropy {of two qubit reduced density matrix} $\widetilde{M_2}(\rho_2)$ (in the inset) and its first derivative with respect to the ratio of the transverse field strength to the nearest neighbor interaction strength $\frac{d\left(\widetilde{M_2}(\rho_2)\right)}{d(\gamma/J_1)}$ as a function of $\gamma/J_1$ for the transverse field Ising model.}
     \label{TFIM_M_2_two}
 \end{figure}

The zero frustration limit ($J_2=0$) of the TANNNI model is the transverse field Ising model (TFIM). We know the quantum phase transition in the TFIM occurs at $\gamma/J_1=1$. However, just like low frustration regime TANNNI model, the critical transition point of $\widetilde{M_2}$ of the TFIM is also far away from the quantum phase transition point. We plot the behavior of $\widetilde{M_2}(\rho_2)$ and its first derivative for TFIM in Fig.~\ref{TFIM_M_2_two}, where the position of the peak in the $\gamma/J_1$ axis is observed to be far away from the quantum phase transition point {\ab \(\gamma/J_1=1\)} and it is closer to the low frustration regime critical transition point of $\widetilde{M_2}$ of the TANNNI model. 

Note that, the transverse-field Ising model (TFIM) is exactly solvable via a sequence of transformations that map the interacting spin system to a quadratic fermionic Hamiltonian~\cite{Lieb2004}. 
The Hamiltonian of the one-dimensional TFIM is given by
\begin{equation}
H_{\mathrm{TFIM}} = -J_1 \sum_{i=1}^{N} \sigma_i^{z}\sigma_{i+1}^{z} - \gamma \sum_{i=1}^{N} \sigma_i^{x},
\end{equation}
where $J_1$ denotes the nearest-neighbor spin-spin coupling strength, $\gamma$ represents the transverse magnetic field, $\sigma_i^{\alpha}$ $(\alpha = x,y,z)$ are the Pauli operators acting on the $i$-th lattice site, and $N$ is the total number of spins. Throughout this work, periodic boundary conditions are assumed such that $\sigma_{N+1}^{\alpha} = \sigma_{1}^{\alpha}$.
After performing the Jordan–Wigner (JW) transformation and Fourier transform, the Hamiltonian takes the quadratic fermionic form
\begin{equation} \label{Fermionic_TFIM}
\begin{split}
H_{\mathrm{TFIM}} 
= \sum_k 
\Big[
2\left(\gamma - J_1 \cos k\right) c_k^\dagger c_k \\
+ i\, 2J_1 \sin k \left( c_{-k}^\dagger c_k^\dagger + c_{-k} c_k \right)
\Big] + \text{const.}
\end{split}
\end{equation}
Without loss of generality, we set $J_1=1$ in the following. After obtaining the quadratic fermionic Hamiltonian~\eqref{Fermionic_TFIM}, we rewrite it in Bogoliubov–de-Gennes (BdG) form in momentum space. For each momentum mode $k$, the quadratic Hamiltonian can be written using the Nambu spinor
\(
\Psi_k =
\begin{pmatrix}
c_k \\
c_{-k}^\dagger
\end{pmatrix},
\)
as
\begin{equation}
H_{{\ab \mathrm{TFIM}}} = \frac{1}{2}\sum_k 
\Psi_k^\dagger \,
\mathcal{H}_{\mathrm{BdG}}(k)
\, \Psi_k
+ \text{const.},
\end{equation}
with the $2\times2$ BdG Hamiltonian for {\ab any} $k$ given by
\begin{equation}
\mathcal{H}_{\mathrm{BdG}}(k)
=
\begin{pmatrix}
\displaystyle 2\left(\gamma - \cos k\right) & \displaystyle 2 i \sin k \\
\displaystyle -2 i \sin k & \displaystyle -2\left(\gamma - \cos k\right)
\end{pmatrix}.
\end{equation}

We use periodic boundary conditions, with allow momenta 
$k=\frac{2\pi m}{N}$, $m=0,1,\ldots,(N-1)$.
For each momentum mode we construct the corresponding $2\times2$ BdG Hamiltonian and diagonalize it independently. After diagonalizing the BdG Hamiltonians, the fermionic two-point correlation functions are obtained in momentum space and subsequently Fourier transformed to real space. The resulting correlation matrix completely characterizes the quadratic ground state and is used to construct the reduced two-qubit density matrix, from which the purity-corrected stabilizer R\'enyi entropy is computed. A detailed description of the procedure is provided in Appendix~\ref{AA}.
The $N=400$ plot of Fig.~\ref{TFIM_M_2_two} is obtained using the method discussed above. The same method is used to study the one dimensional quantum compass model in Sec.~\ref{III}.

\begin{figure}
    \centering
       \includegraphics[width=0.99\linewidth]{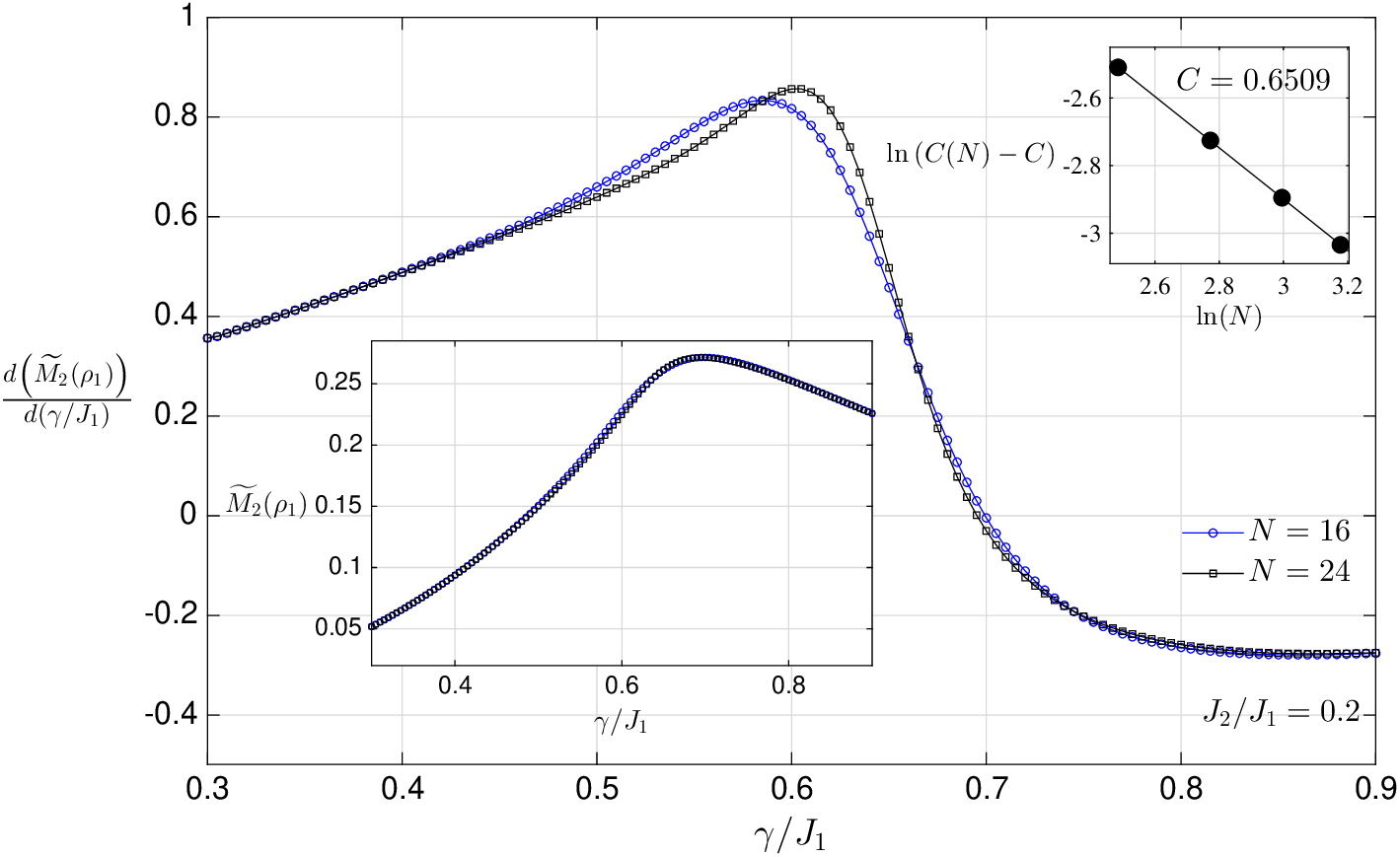}
   (a) \includegraphics[width=1\linewidth]{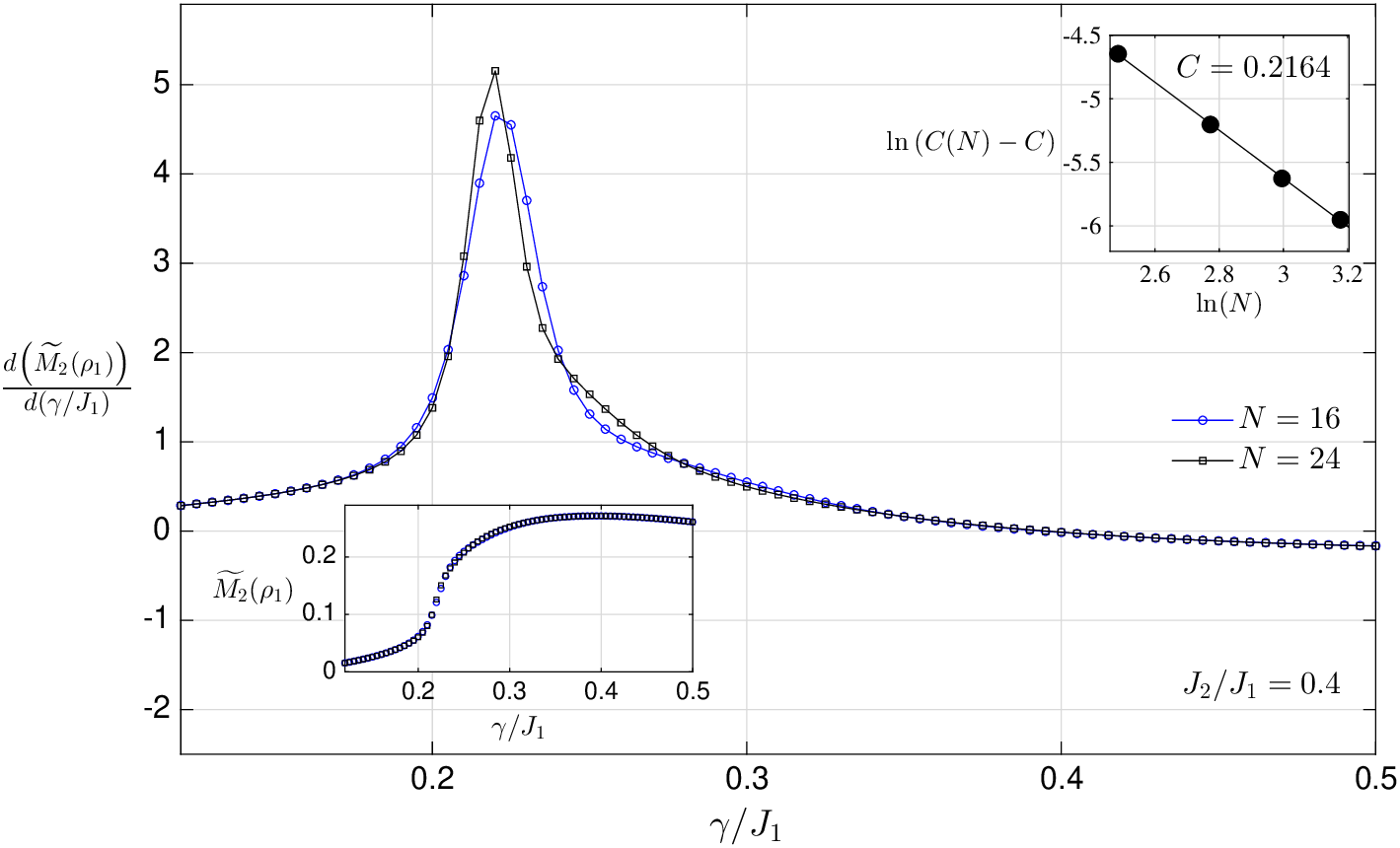}
   (b) 
    \caption{Purity corrected stabilizer Rényi entropy (in left inset) {{of one qubit reduced density matrix}} ($\widetilde{M_2}(\rho_1)$) and its first derivative with respect to the ratio of the transverse field strength to the nearest neighbor interaction strength, $\frac{d\left(\widetilde{M_2}(\rho_1)\right)}{d(\gamma/J_1)}$, plotted as functions of $\gamma/J_1$ for different frustration parameters: (a) $J_2/J_1 = 0.2$, (b) $J_2/J_1 = 0.4$. Each panel shows results for two system sizes, $N = 16$ (blue curves) and $N = 24$ (black curves). The right inset panels display the finite-size scaling behavior of the maxima in $\frac{d\left(\widetilde{M_2}(\rho_1)\right)}{d(\gamma/J_1)}$. The critical points in the thermodynamic-limit are estimated from the finite-size scaling behavior. The estimated thermodynamic-limit critical points ($C$) corresponding to each frustration parameter are indicated in the respective right inset panels.
}
    \label{one_left_ANNNI_sre-re_1}
\end{figure}

\begin{figure}
    \centering
       \includegraphics[width=0.99\linewidth]{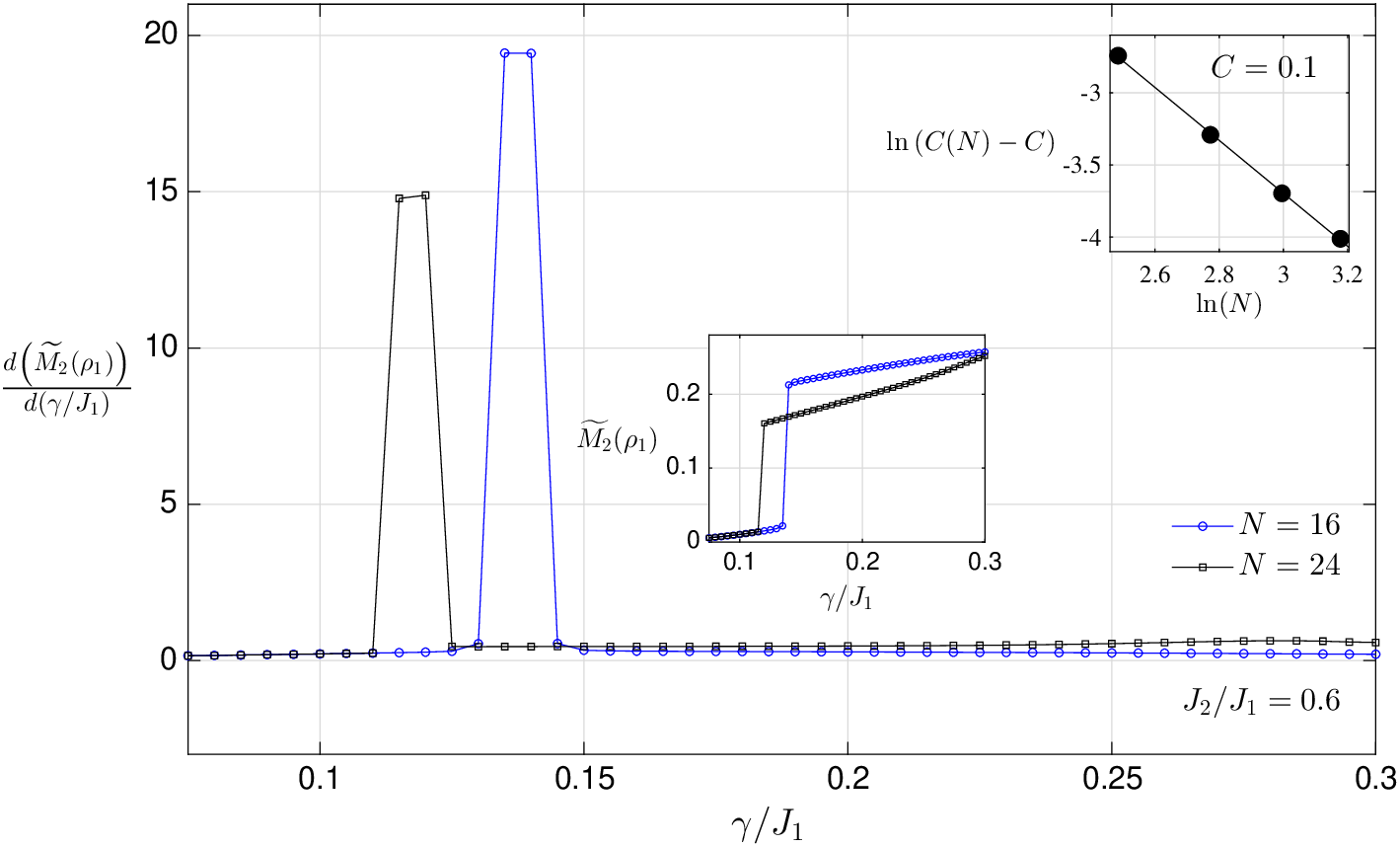}
   (a) \includegraphics[width=1\linewidth]{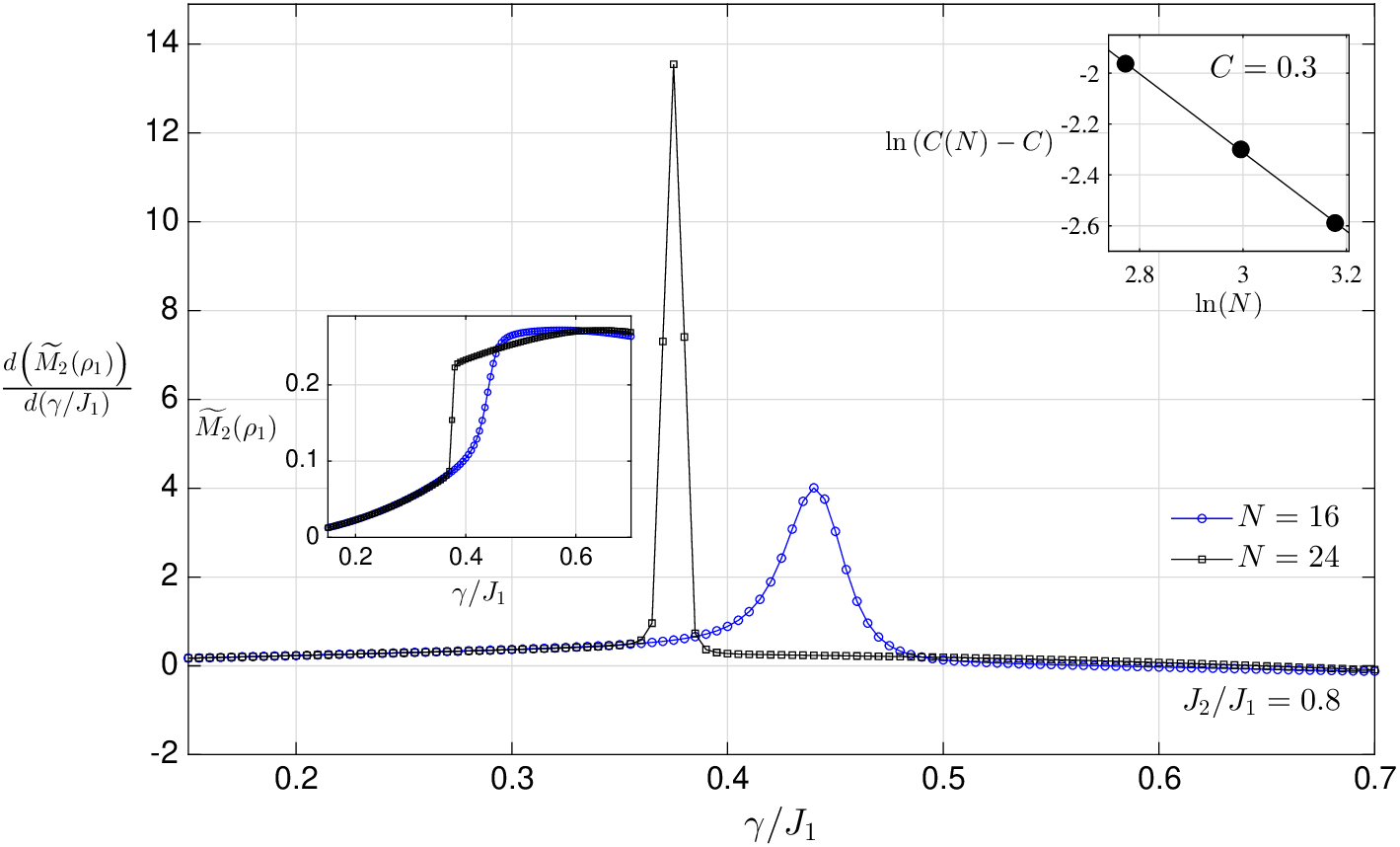}
   (b) 
    \caption{Purity corrected stabilizer Rényi entropy (in left inset) {{of one qubit reduced density matrix}} ($\widetilde{M_2}(\rho_1)$) and its first derivative with respect to the ratio of the transverse field strength to the nearest neighbor interaction strength, $\frac{d\left(\widetilde{M_2}(\rho_1)\right)}{d(\gamma/J_1)}$, plotted as functions of $\gamma/J_1$ for different frustration parameters: (a) $J_2/J_1 = 0.6$, (b) $J_2/J_1 = 0.8$. Each panel shows results for two system sizes, $N = 16$ (blue curves) and $N = 24$ (black curves). The right inset panels display the finite-size scaling behavior of the maxima in $\frac{d\left(\widetilde{M_2}(\rho_1)\right)}{d(\gamma/J_1)}$, demonstrating the convergence of the finite-size critical points $C(N)$ toward the known thermodynamic-limit values with increasing system size. The known quantum critical points ($C$) corresponding to each frustration parameter are indicated in the respective right inset panels.
}
    \label{one_right_ANNNI_sre-re_2}
\end{figure}

We next study the purity-corrected SRE, \(\widetilde{M_2}\), of the one-qubit reduced density matrix \(\rho_1\) for the TANNNI model. We observe that, although the numerical values of \(\widetilde{M_2}(\rho_1)\) differ from those of \(\widetilde{M_2}(\rho_2)\), the scaled positions of the critical transitions are approximately identical in both the one- and two-qubit cases. In particular, the peaks of \(\frac{d\left(\widetilde{M_2}(\rho_1)\right)}{d\left(\gamma/J_1\right)}\) occur at nearly the same values of \(\gamma/J_1\) as the peaks of \(\frac{d\left(\widetilde{M_2}(\rho_2)\right)}{d\left(\gamma/J_1\right)}\), indicating convergence to the same critical points, {\ab demonstrating} that both one-{\ab qubit} and two-qubit reduced states capture the same underlying critical behavior. A few one-qubit results are shown in Fig.~\ref{one_left_ANNNI_sre-re_1} for the \(J_2/J_1 < 0.5\) regime and in Fig.~\ref{one_right_ANNNI_sre-re_2} for the \(J_2/J_1 > 0.5\) regime.

\begin{figure}
    \centering
       \includegraphics[width=0.99\linewidth]{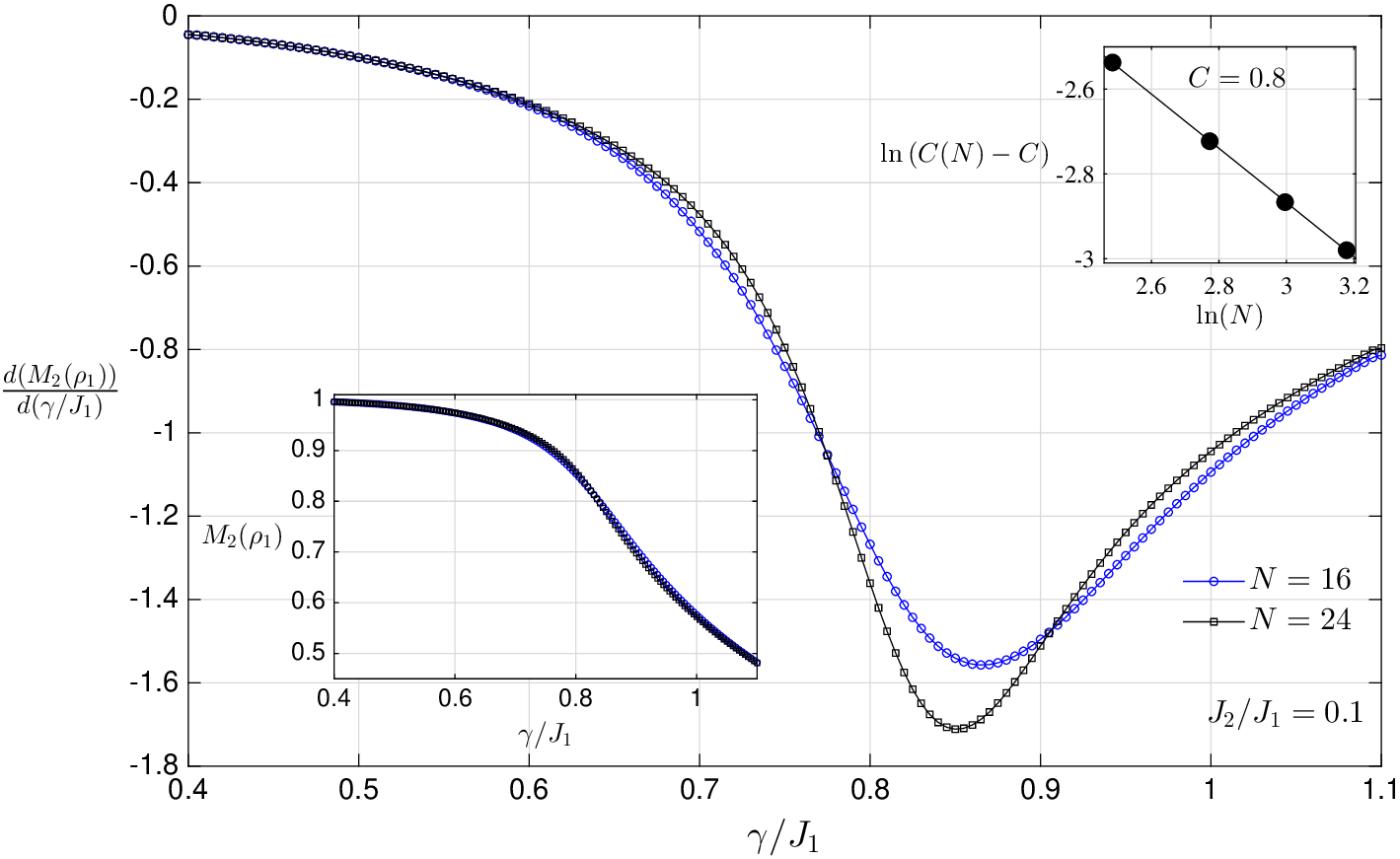}
   (a) \includegraphics[width=1\linewidth]{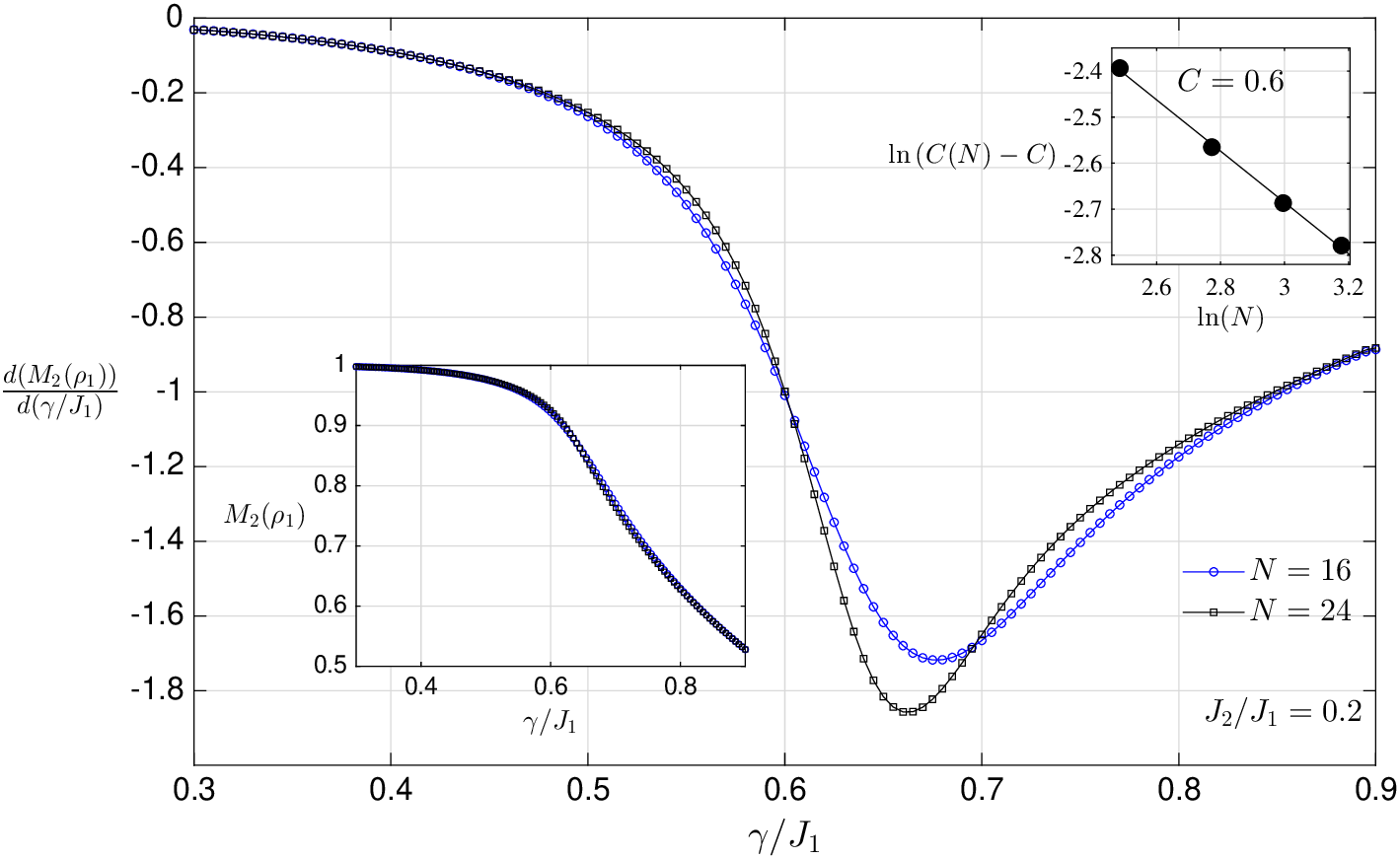}
   (b) 
    \caption{Purity uncorrected stabilizer Rényi entropy (in left inset) {{of one qubit reduced density matrix}} (${M_2}(\rho_1)$) and its first derivative with respect to the ratio of the transverse field strength to the nearest neighbor interaction strength, $\frac{d\left({M_2}(\rho_1)\right)}{d(\gamma/J_1)}$, plotted as functions of $\gamma/J_1$ for different frustration parameters: (a) $J_2/J_1 = 0.1$, (b) $J_2/J_1 = 0.2$. Each panel shows results for two system sizes, $N = 16$ (blue curves) and $N = 24$ (black curves). The right inset panels display the finite-size scaling behavior of the minima in $\frac{d\left({M_2}(\rho_1)\right)}{d(\gamma/J_1)}$, demonstrating the convergence of the finite-size critical points $C(N)$ toward the known thermodynamic-limit values with increasing system size. The known quantum critical points ($C$) corresponding to each frustration parameter are indicated in the respective right inset panels.
}
    \label{left_raw_1}
\end{figure}

\begin{figure}
    \centering
       \includegraphics[width=0.99\linewidth]{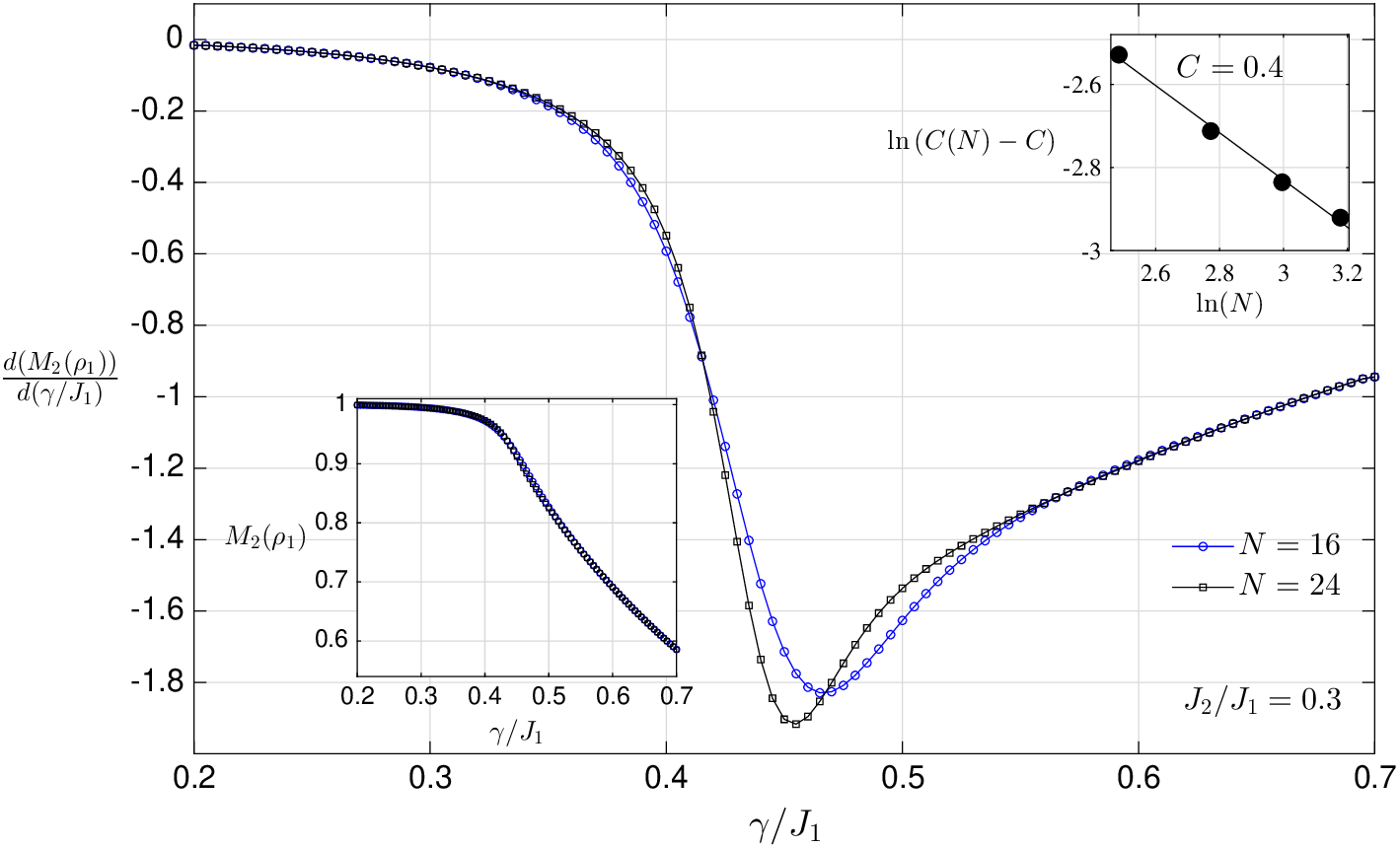}
   (a) \includegraphics[width=1\linewidth]{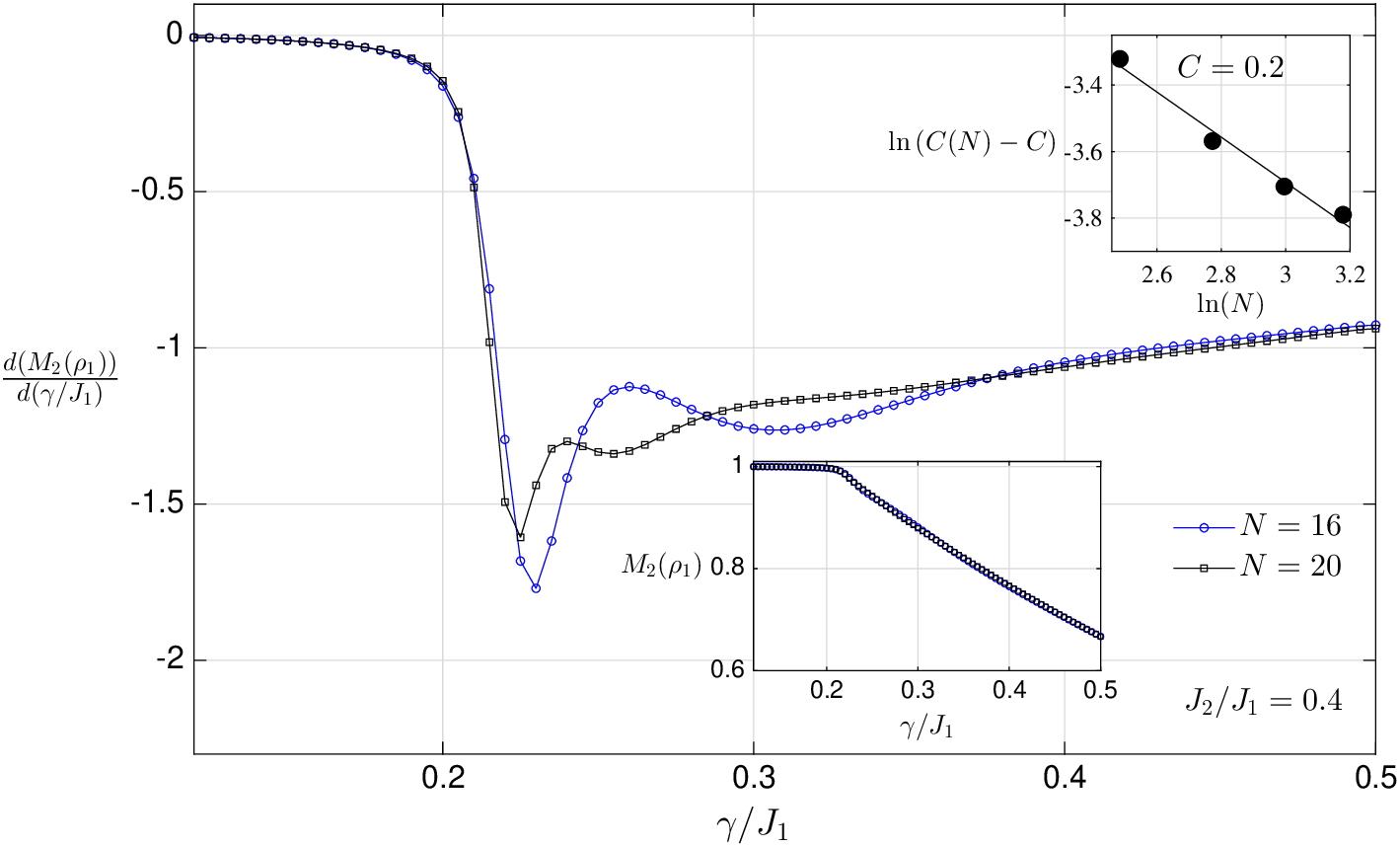}
   (b) 
    \caption{Purity uncorrected stabilizer Rényi entropy (in left inset) {{of one qubit reduced density matrix}} (${M_2}(\rho_1)$) and its first derivative with respect to the ratio of the transverse field strength to the nearest neighbor interaction strength, $\frac{d\left({M_2}(\rho_1)\right)}{d(\gamma/J_1)}$, plotted as functions of $\gamma/J_1$ for different frustration parameters: (a) $J_2/J_1 = 0.3$, (b) $J_2/J_1 = 0.4$. Each panel shows results for two system sizes, $N = 16$ (blue curves) and $N = 24$ (black curves). The right inset panels display the finite-size scaling behavior of the minima in $\frac{d\left({M_2}(\rho_1)\right)}{d(\gamma/J_1)}$, demonstrating the convergence of the finite-size critical points $C(N)$ toward the known thermodynamic-limit values with increasing system size. The known quantum critical points ($C$) corresponding to each frustration parameter are indicated in the respective right inset panels.
}
    \label{left_raw_2}
\end{figure}

\begin{figure}
    \centering
       \includegraphics[width=0.99\linewidth]{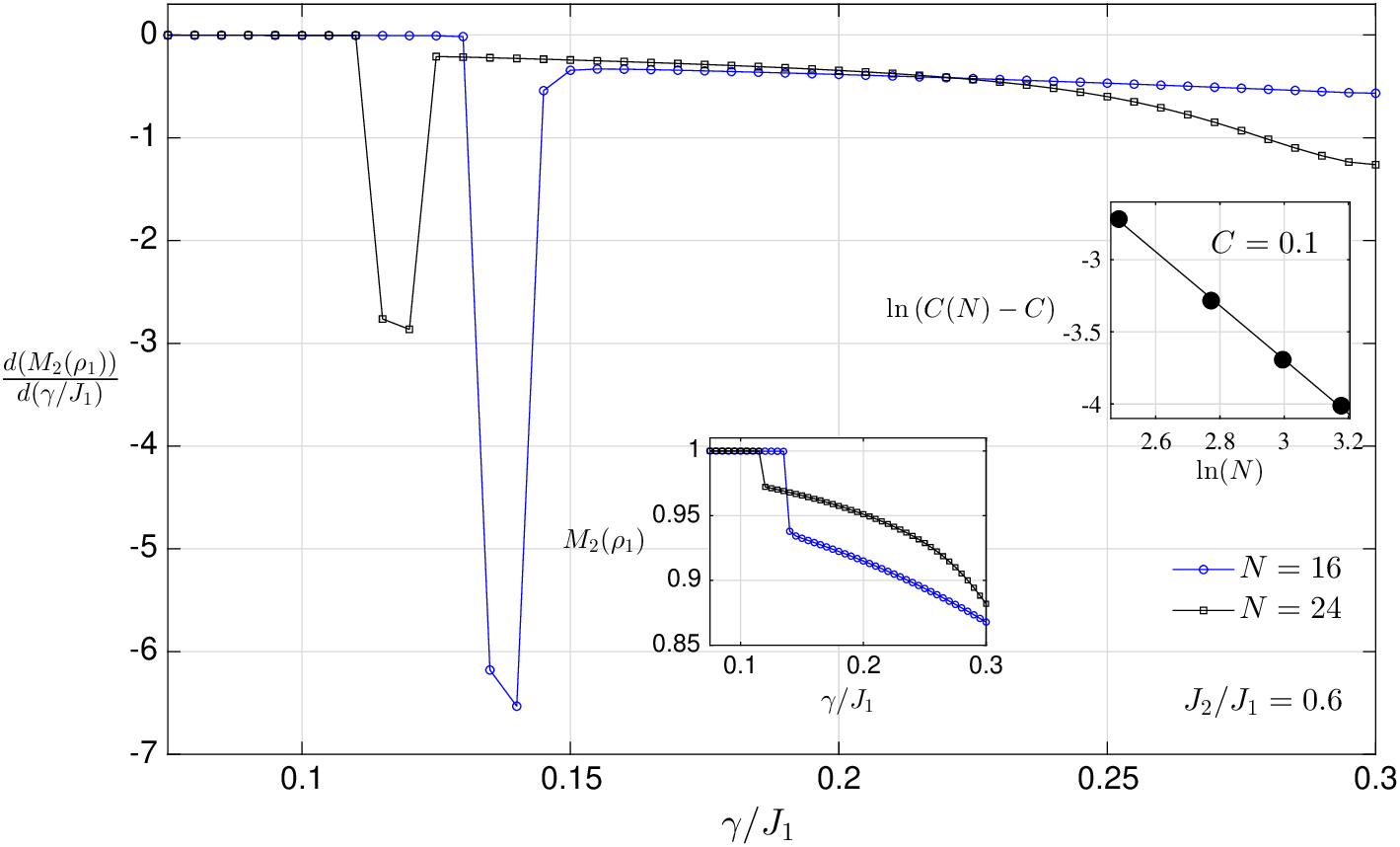}
   (a) \includegraphics[width=1\linewidth]{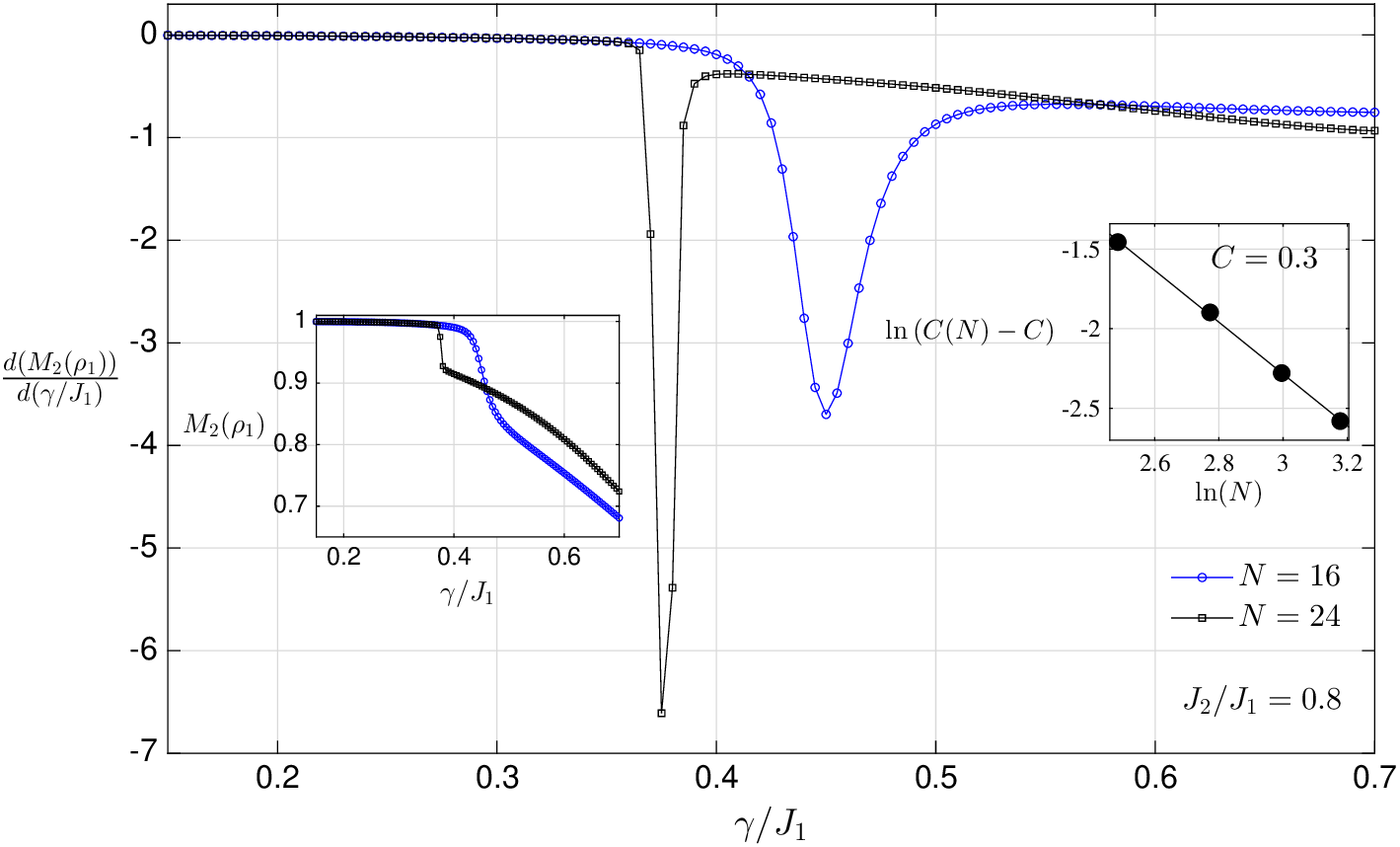}
   (b) 
    \caption{Purity uncorrected stabilizer Rényi entropy (in left inset) {{of one qubit reduced density matrix}} (${M_2}(\rho_1)$) and its first derivative with respect to the ratio of the transverse field strength to the nearest neighbor interaction strength, $\frac{d\left({M_2}(\rho_1)\right)}{d(\gamma/J_1)}$, plotted as functions of $\gamma/J_1$ for different frustration parameters: (a) $J_2/J_1 = 0.6$, (b) $J_2/J_1 = 0.8$. Each panel shows results for two system sizes, $N = 16$ (blue curves) and $N = 24$ (black curves). The right inset panels display the finite-size scaling behavior of the minima in $\frac{d\left({M_2}(\rho_1)\right)}{d(\gamma/J_1)}$, demonstrating the convergence of the finite-size critical points $C(N)$ toward the known thermodynamic-limit values with increasing system size. The known quantum critical points ($C$) corresponding to each frustration parameter are indicated in the respective right inset panels.
}
    \label{right_raw}
\end{figure}

Another interesting observation concerns the behavior of the raw (i.e., purity-uncorrected) stabilizer R\'enyi entropy \(M_2\) in the TANNNI model. We compute \(M_2\) for the one-qubit reduced density matrix and observe a pronounced dip in its first derivative, \(\frac{d\left(M_2(\rho_1)\right)}{d\left(\gamma/J_1\right)}\), as a function of \(\gamma/J_1\). In the high-frustration regime (\(J_2/J_1 > 0.5\)), the positions of these dips closely coincide with the peak positions of the derivative of the purity-corrected SRE, \(\frac{d\left(\widetilde{M_2}(\rho_1)\right)}{d\left(\gamma/J_1\right)}\), indicating consistency between the two quantities in identifying the transition points. In contrast, in the low-frustration regime (\(J_2/J_1 < 0.5\)), the dip positions deviate from {\ab peak} positions of the purity corrected SRE and instead approach the previously known quantum phase transition points, indicating that in the low frustration region, the uncorrected SRE carries signatures of criticality better. We show the corresponding results in Fig.~\ref{left_raw_1} and Fig.~\ref{left_raw_2} for the low-frustration regime (\(J_2/J_1<0.5\)), and in Fig.~\ref{right_raw} for the high-frustration regime (\(J_2/J_1>0.5\)). The transition points extracted from the raw \(M_2\) are also summarized in the phase diagram of Fig.~\ref{schematic_2}, where they are represented by blue triangular markers.

\section{Quantum Compass Model}\label{III}

The one-dimensional quantum compass model (1-D QCM) is a paradigmatic example of a frustrated spin system with bond-directional interactions. Unlike conventional spin models where the same spin component interacts along all bonds, the compass model features alternating couplings of different spin components, leading to intrinsic frustration and nontrivial ground-state properties. The model originated in orbital physics as an effective description of directional orbital interactions in strongly correlated materials.~\cite{Kugel1982,RevModPhys.87.1}.

The Hamiltonian of the one-dimensional quantum compass model is given by
\begin{equation}
H_{\mathrm{QCM}} = - \sum_{i=1}^{N/2} 
\left( J_x \, \sigma_{2i-1}^{x}\sigma_{2i}^{x} 
+ J_z \, \sigma_{2i}^{z}\sigma_{2i+1}^{z} \right),
\end{equation}
where $J_x$ and $J_z$ denote the coupling strengths along alternating bonds.

A central property of the quantum compass model is the quantum phase transition occurring at the isotropic point $J_x/J_z = 1$. For $J_x > J_z$, the system exhibits dominant ordering in the $\sigma^x$ component, whereas for $J_z > J_x$, the ordering is primarily along the $\sigma^z$ direction. At the isotropic point $J_x = J_z$, the system undergoes a first-order quantum phase transition characterized by a discontinuous change in the ground-state properties~\cite{PhysRevB.75.134415,PhysRevB.79.224424}.

To analyze the 1-D QCM, we follow the exact solution procedure outlined in Ref.~\cite{PhysRevB.79.224424}. {\ab After performing the Jordan–Wigner transformation and Fourier transform, the Hamiltonian
takes the quadratic fermionic form}
\begin{equation} \label{Fermionic_QCM}
\begin{split}
H_{QCM} = J_z \sum_k 
[
2\left(\frac{J_x}{J_z} - \cos k\right) c_k^\dagger c_k \\
+ i \sin k \left( c_{-k}^\dagger c_k^\dagger + c_{-k} c_k \right)
] + \text{const.}
\end{split}
\end{equation}
After obtaining the quadratic fermionic Hamiltonian {\ab given by Eq.~(\ref{Fermionic_QCM})} for 1-D QCM (which is similar to the {\ab one for} TFIM), we rewrite it in BdG form in momentum space. 
For each momentum mode $k$, the quadratic Hamiltonian can be written in Bogoliubov--de Gennes (BdG) form using the Nambu spinor
\(
\Psi_k=\begin{pmatrix} c_k \\ c_{-k}^\dagger \end{pmatrix},
\)
as
\begin{equation}
H_{QCM}=\frac{1}{2}\sum_k \Psi_k^\dagger \, \mathcal{H}_{\mathrm{BdG}}(k)\, \Psi_k + \text{const.},
\end{equation}
with the $2\times 2$ BdG Hamiltonian for {\ab any} $k$ given by
\begin{equation}
\mathcal{H}_{\mathrm{BdG}}(k)
=
\begin{pmatrix}
\displaystyle \left(2 J_x-\cos k\right) & i\sin k \\
- i\sin k & \displaystyle -\left(2 J_x -\cos k\right)
\end{pmatrix},
\end{equation}
where, without loss of generality we set \(J_z=1\).
For each momentum mode $k = 0, \pm \frac{2\pi}{(N/2)}, \pm 2\frac{2\pi}{(N/2)}, \ldots, \pi$ or $
k = \pm \frac{1}{2}\frac{2\pi}{(N/2)}, \pm \frac{3}{2}\frac{2\pi}{(N/2)}, \ldots,
\pm \frac{1}{2}((N/2)-1)\frac{2\pi}{(N/2)}.
$~\cite{PhysRevB.75.134415,PhysRevB.79.224424}, we construct the corresponding $2\times2$ BdG Hamiltonian and diagonalize it independently. 

From the diagonalized BdG Hamiltonians, we compute the fermionic two-point correlation functions in momentum space and transform them back to real space. 
Using the real-space correlation matrix, we evaluate the reduced density matrix. The step by step procedure are given in Appendix~\ref{AA}.

In Fig.~\ref{F_QCM} and Fig.~\ref{F_QCM_one}, we plot the purity-corrected SRE of the two-qubit reduced density matrix, \(\widetilde{M_2}(\rho_2)\), and the one-qubit reduced density matrix, \(\widetilde{M_2}(\rho_1)\), respectively, as functions of the coupling ratio \(J_x/J_z\). In both cases, \(\widetilde{M_2}\) exhibits a clear change in curvature in the vicinity of the quantum critical point. Correspondingly, the first derivative of \(\widetilde{M_2}\) with respect to \(J_x/J_z\) develops a pronounced dip near the same location. These features provide a clear signature of the phase transition and demonstrate that the purity-corrected SRE serves as an effective local probe of quantum criticality in the one-dimensional quantum compass model.

 \begin{figure}
     \centering
     \includegraphics[width=0.99\linewidth]{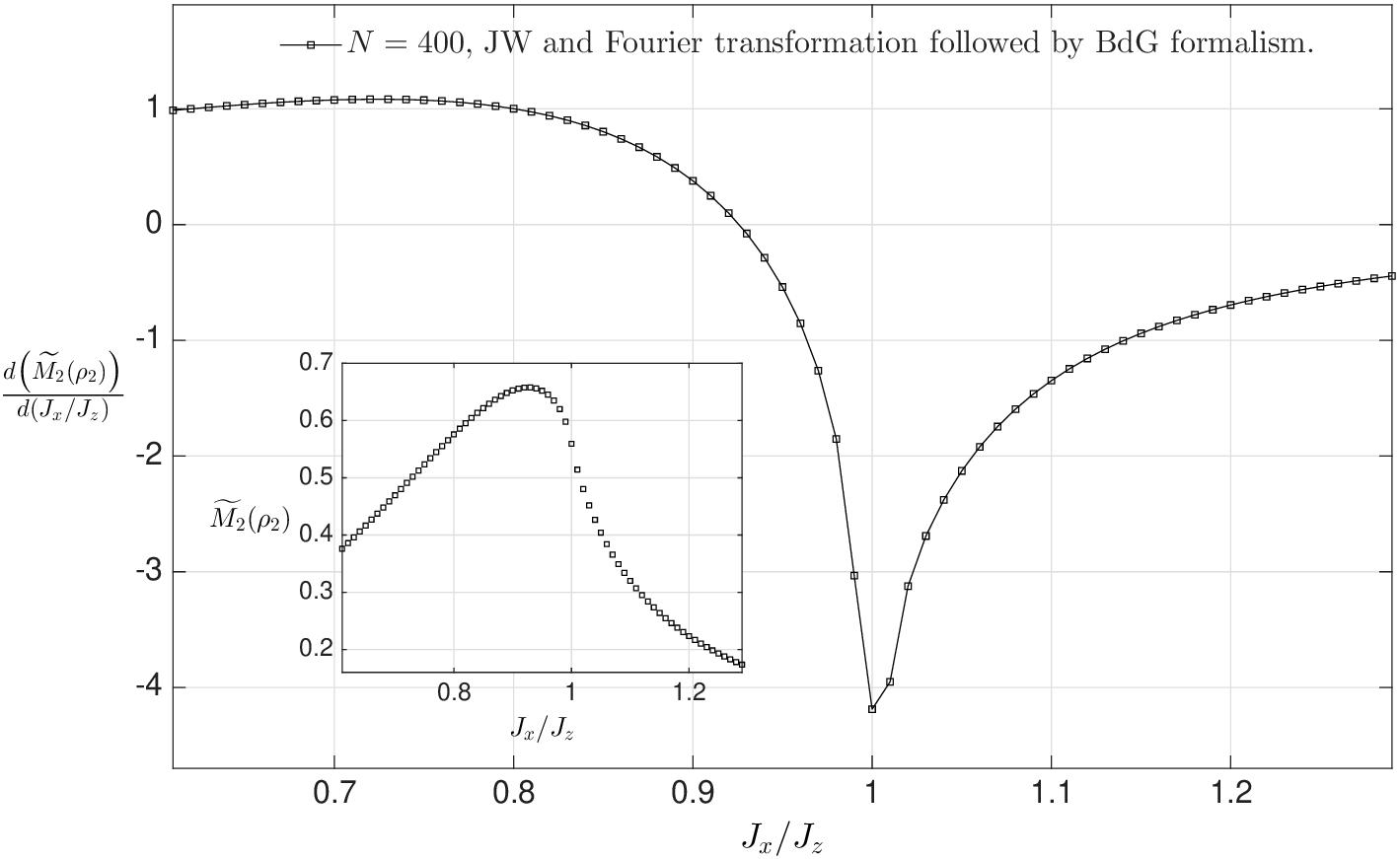}
     \caption{Purity corrected stabilizer Rényi entropy {of two qubit reduced density matrix} $\widetilde{M_2}(\rho_2)$ (in the inset) and its first derivative with respect to the coupling ratio $\frac{d\left(\widetilde{M_2}(\rho_2)\right)}{d(J_x/J_z)}$ as a function of the coupling ratio $J_x/J_z$ for the one-dimensional quantum compass model. The plot shows that a pronounced dip in $\frac{d\left(\widetilde{M_2}(\rho_2)\right)}{d(J_x/J_z)}$ appears at the quantum phase transition point $J_x/J_z = 1$.}
     \label{F_QCM}
 \end{figure}

 \begin{figure}
     \centering
     \includegraphics[width=0.99\linewidth]{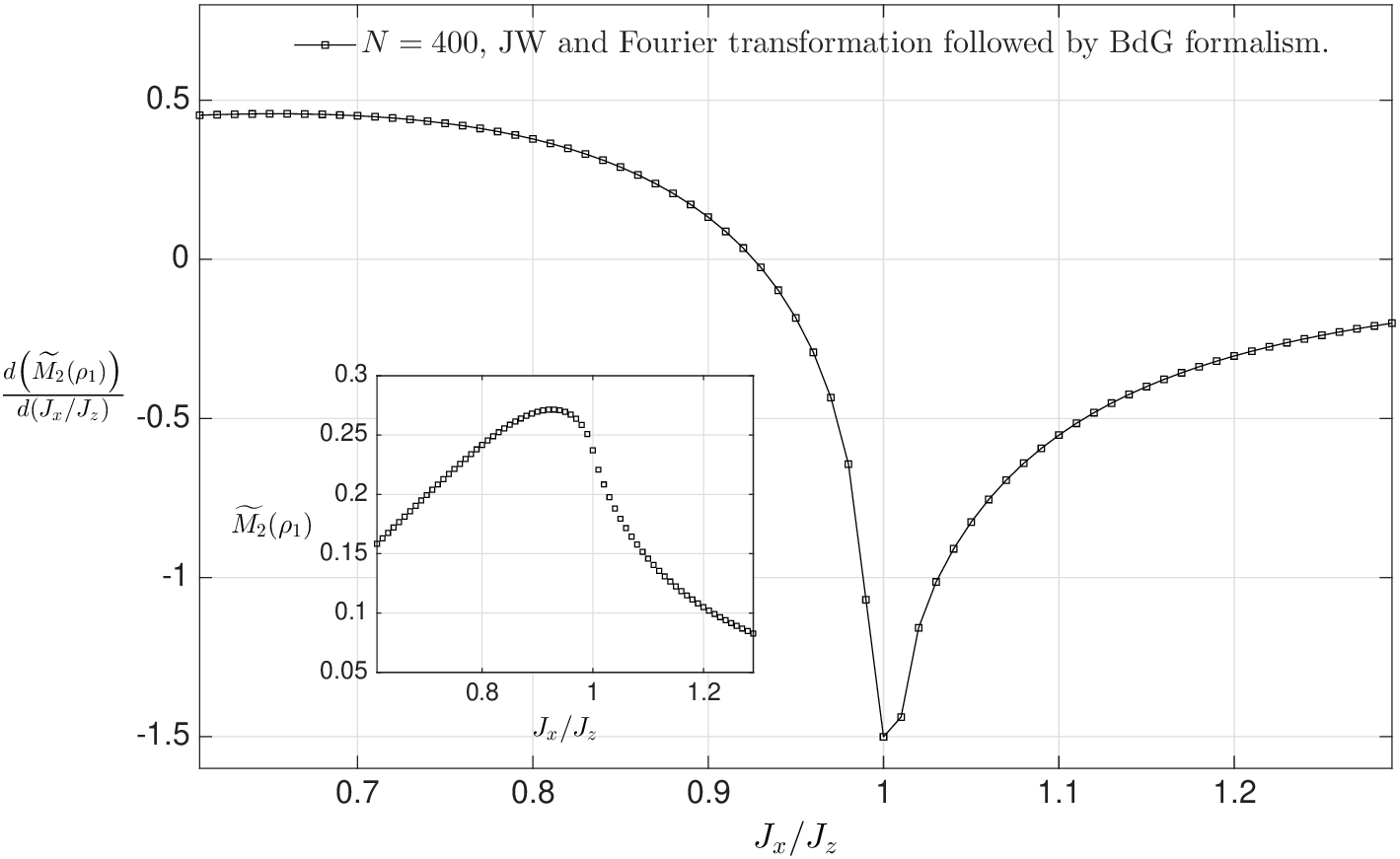}
     \caption{{Purity corrected stabilizer Rényi entropy {of one qubit reduced density matrix} $\widetilde{M_2}(\rho_1)$ (in the inset) and its first derivative with respect to the coupling ratio $\frac{d\left(\widetilde{M_2}(\rho_1)\right)}{d(J_x/J_z)}$ as a function of the coupling ratio $J_x/J_z$ for the one-dimensional quantum compass model. The plot shows that a pronounced dip in $\frac{d\left(\widetilde{M_2}(\rho_1)\right)}{d(J_x/J_z)}$ appears at the quantum phase transition point $J_x/J_z = 1$.}}
     \label{F_QCM_one}
 \end{figure}

\section{Conclusion}\label{IV}

In this work, we have investigated the purity-corrected stabilizer R\'enyi entropy (SRE), a quantifier associated with non-stabilizer resources or quantum magic, as a probe of quantum phase transitions in two paradigmatic one-dimensional spin systems: the transverse axial next-nearest-neighbor Ising (TANNNI) model and the quantum compass model (QCM). By evaluating the purity-corrected SRE of reduced density matrices derived from the many-body ground state, we demonstrated that local magic-based measures can capture signatures of critical behavior.

For the TANNNI model, in the high-frustration regime \(J_2/J_1>0.5\), the purity-corrected SRE successfully detects the antiphase--floating phase transition, and finite-size scaling shows reliable convergence toward the known critical points. In the low-frustration regime \(J_2/J_1<0.5\), however, the transition points obtained from the purity-corrected SRE do not coincide with the known ferromagnetic--paramagnetic phase boundary. Interestingly, the transitions obtained from the raw (purity-uncorrected) SRE agree well with the known quantum critical points in this regime. Since the mixedness of the reduced density matrices originates from entanglement between the reduced subsystem and the traced-out remainder of the system, the contribution of entanglement appears to play an important role in capturing the phase boundaries of the TANNNI model. Consequently, in the low-frustration regime, the purity-uncorrected SRE provides a better indicator of the transition points.

For the one-dimensional quantum compass model, we employed the Jordan--Wigner transformation, Fourier transformation, and Bogoliubov--de-Gennes formalism to obtain the ground-state correlation functions by exploiting the close relation of the model to the transverse field Ising model. The purity-corrected SRE exhibits a clear signature near the isotropic point \(J_x/J_z=1\), where the system undergoes a first-order quantum phase transition. In particular, the sharp dip in the derivative of the purity-corrected SRE provides a direct indication of the transition, demonstrating that quantum magic can effectively probe quantum criticality in the 1-D QCM.

Our results demonstrate that the stabilizer R\'enyi entropy of reduced density matrices (both purity-corrected and uncorrected) serves as a complementary quantum-correlation measure for probing many-body critical phenomena. The behavior of the reduced-state SRE near the transition points indicates that quantum phase transitions in the spin models considered here are accompanied by a substantial restructuring of non-stabilizer resources. The reduced-density-matrix approach adopted here avoids the exponential complexity associated with evaluating SRE for full many-body states while still retaining the essential signatures of criticality. These results open avenues for further investigations of quantum magic in other spin models, disordered systems, and dynamical quantum phase transitions~\cite{Heyl_2018}, and provide additional insight into the interplay between computational quantum resources and many-body quantum phenomena.

\acknowledgements
GB gratefully acknowledges Shao-Hua Hu for valuable discussions and important suggestions. GB is supported by National Science and Technology Council, Taiwan under the Grant No. 114-2811-M-032-006. SS acknowledges support from the Council of Scientific and Industrial Research through the grant (award number: 09/1336(11432)/2021-EMR-I). JYW is supported by National Science and Technology Council (NSTC), Taiwan under Grants No. 114-2119-M-008-008, 113-2119-M-008-010, 112-2112-M-032-008-MY3, and 111-2923-M-032-002-MY5. 

\appendix
\section{} \label{AA}

In the following, we outline the procedure for computing the reduced two-qubit density matrix and stabilizer R\'enyi entropy (SRE) starting from the Bogoliubov--de Gennes (BdG) Hamiltonian. Here $N'$ denotes the number of fermionic modes: $N'=N$ for the TFIM and $N'=N/2$ for the QCM, where $N$ is the number of spin sites.

From the diagonalized BdG Hamiltonian, we take the eigenvector $\begin{bmatrix}
    u_k\\ v_k
\end{bmatrix}$ corresponding to minimum eigenvalue, and compute the fermionic two-point correlators in momentum space:
\begin{equation}
\langle c_k^\dagger c_k \rangle = \abs{v_k}^2,
\qquad
\langle c_k c_{-k} \rangle = u_k v_k.
\end{equation}

Transforming back to real space using
\[
c_j = \frac{1}{\sqrt{N'}} \sum_k e^{ikj} c_k,
\]
we obtain
\begin{equation}
\langle c_i^\dagger c_j \rangle
=
\frac{1}{N'} \sum_k e^{ik(i-j)} \abs{v_k}^2,
\end{equation}
\begin{equation}
\langle c_i c_j \rangle
=
\frac{1}{N'} \sum_k e^{ik(i-j)} u_k v_k.
\end{equation}

These expectation values define the correlation matrix
\[
G_{ij} = \langle c_i^\dagger c_j \rangle,
\qquad
F_{ij} = \langle c_i c_j \rangle.
\]

Since the Hamiltonian is quadratic, Wick’s theorem applies and all higher-order correlators can be expressed in terms of $G_{ij}$ and $F_{ij}$~\cite{Peschel_2009,Ingo_Peschel_2003}.

To compute the stabilizer R\'enyi entropy (SRE) from the two-site correlation matrix, we proceed as follows.

Starting from the $4\times4$ fermionic sub-correlation matrix~\cite{Ingo_Peschel_2003}
\[
C =
\begin{pmatrix}
G_{12} & F_{12}^\dagger \\
F_{12} & \mathbb{I}_2 - G_{12}^{T}
\end{pmatrix},
\]
we first transform it to the Majorana basis~\cite{PhysRevLett.90.227902,10.5555/2011572.2011576}. For $N=2$ fermionic modes, we define the Majorana transformation matrix
\[
W =
\begin{pmatrix}
\mathbb{I}_2 & \mathbb{I}_2 \\
- i \mathbb{I}_2 & i \mathbb{I}_2
\end{pmatrix}.
\]
The Majorana covariance matrix is then obtained as
\begin{equation}
M = W C W^\dagger,
\end{equation}
and the antisymmetric covariance matrix
\begin{equation}
\Gamma = \frac{i}{2}(M - M^T).
\end{equation}

From the Majorana covariance matrix, we extract the expectation values of the relevant Pauli operators. Using the correspondence between Majorana bilinears and spin operators under the Jordan--Wigner transformation, the single-site expectations are
\begin{equation}
\langle Z_1 \rangle = \Gamma_{1,2}, 
\qquad
\langle Z_2 \rangle = \Gamma_{3,4},
\end{equation}
while the two-site transverse correlators are
\begin{align}
\langle XX \rangle &= \Gamma_{2,3}, \\
\langle YY \rangle &= \Gamma_{1,4}, \\
\langle XY \rangle &= \Gamma_{2,4}, \\
\langle YX \rangle &= -\Gamma_{1,3}.
\end{align}
The longitudinal correlator is obtained from Wick's theorem as
\begin{equation}
\langle ZZ \rangle
=
- \Big(
\Gamma_{1,2}\Gamma_{3,4}
- \Gamma_{1,3}\Gamma_{2,4}
+ \Gamma_{1,4}\Gamma_{2,3}
\Big).
\end{equation}

Using these correlation functions, the reduced two-qubit density matrix \(\rho_2\) is reconstructed through the Pauli expansion
\begin{equation}
\rho_2
=
\frac{1}{4}
\sum_{\alpha,\beta \in \{I,X,Y,Z\}}
\langle \sigma^\alpha \otimes \sigma^\beta \rangle
\,\sigma^\alpha \otimes \sigma^\beta .
\end{equation}

The second-order stabilizer R\'enyi entropy is then computed from the reduced density matrix as~\cite{PhysRevLett.128.050402,Haug2023stabilizerentropies,PhysRevA.110.L040403}
\begin{equation}
M_2(\rho_2)
=
- \log_2
\left(
\frac{1}{4}
\sum_{\alpha,\beta}
\left[
\mathrm{Tr}
\left(
\rho_2 \,
\sigma^\alpha \otimes \sigma^\beta
\right)
\right]^4
\right).
\end{equation}

Since the reduced density matrix is generally mixed, we also evaluate the purity-corrected stabilizer R\'enyi entropy,
\begin{equation}
\widetilde{M_2}(\rho_2)
=
M_2(\rho_2)
-
S_2(\rho_2),
\end{equation}
where
\begin{equation}
S_2(\rho_2)
=
-\log_2\left(\mathrm{Tr}(\rho_2^2)\right)
\end{equation}
is the second-order R\'enyi entropy of the reduced density matrix.

Thus, both the raw and purity-corrected SRE are obtained directly from the two-site fermionic correlation matrix through the Majorana covariance matrix and the reconstructed reduced density matrix.

\bibliography{References}
\end{document}